\documentclass[
reprint,
onecolumn,
superscriptaddress,
preprintnumbers,
amsmath,amssymb, aps
]{revtex4-2}
\pdfoutput=1

\usepackage[utf8]{inputenc}
\usepackage{graphicx}
\usepackage{amsmath}
\DeclareMathOperator{\csch}{csch}
\usepackage{amssymb}
\usepackage{array}

\usepackage[usenames,dvipsnames]{xcolor}
\usepackage{hyperref}

\begin{document}

\title{Quantum Noise from Massive Lorentz--Violating Gravitons:\\
Interferometer Signatures in Very Special Linear Gravity}

\author{Jorge Alfaro}
\email{jalfaro@uc.cl}
\affiliation{Instituto de Física, Pontificia Universidad Católica de Chile, \\
 	Casilla 306, Santiago 22, Chile.}
 
\author{Alex Soto}
\email{arsoto@ifpan.edu.pl}
\affiliation{Institute of Physics, Polish Academy of Sciences,\\
Aleja Lotników 32/46, PL-02-668 Warsaw, Poland}

\author{Alessandro Santoni}
\email{asantoni@uc.cl}
\affiliation{Institute of Space Sciences and Astronomy, University of Malta, Msida, MSD 2080, Malta}
\affiliation{Department of Physics, University of Malta, Msida, MSD 2080, Malta}

\date{\today}

\begin{abstract}
\noindent In this work we analyse under the Very Special Linear Gravity framework, which introduces a gauge invariant massive graviton with two polarizations, possible signatures of gravitons in a classical gravitational detector. Following the Vernon approach we found dissipative terms consistent with other works and extra terms of order of the graviton mass squared. The model also presents a characteristic feature when a squeezed state is considered. In such case, a contribution that distinguishes the sign of the squeezing phase is present.
\end{abstract}

\maketitle

\section{Introduction}

The first observation of gravitational waves (GWs) by LIGO \cite{LIGOScientific:2016aoc} established interferometers as a tool for testing gravity in the dynamical regime and motivated the study of possible signatures of quantum gravity in detectors. Coupling a detector to quantized metric perturbations induces fluctuations in the separation of its test masses, with correlations that depend on the gravitational field state \cite{PhysRevD.104.046021}. These fluctuations can therefore carry information about the quantization of the field, although identifying a measurable noise contribution and establishing its quantum origin are separate tasks.

In the weak-field regime, quantization of gravity introduces the graviton, a spin-2 particle that is massless in the standard description. Massive spin-2 fields were first studied by Fierz and Pauli \cite{Fierz:1939ix} and later became central to theories of massive gravity. In the Lorentz-invariant Fierz-Pauli theory, a massive graviton carries five physical degrees of freedom, rather than the two of linearized general relativity. Its linear coupling to sources exhibits the van Dam-Veltman-Zakharov discontinuity: the massless limit does not reproduce the predictions of general relativity \cite{vanDam:1970vg,Zakharov:1970cc}. Generic nonlinear extensions can also introduce the Boulware-Deser ghost \cite{Boulware:1972yco}. Nonlinear Vainshtein screening \cite{Vainshtein:1972sx}, relevant in DGP \cite{Dvali:2000hr,Dvali:2000rv} and dRGT gravity \cite{deRham:2010kj}, can recover general-relativistic predictions near sources. Avoiding the ghost is a separate consistency requirement of the nonlinear theory.

These features need not accompany every graviton mass once Lorentz symmetry is relaxed. Very Special Relativity (VSR), introduced in 2006, replaces Lorentz symmetry by a subgroup, here $SIM(2)$ \cite{vsr1}. Developed independently of the graviton-mass problem, VSR also offers a possible origin of neutrino masses without additional right-handed neutrino fields \cite{vsr2}. The subgroup preserves a preferred null direction $n^\mu$ up to rescaling. Ratios such as $n^\mu/(n\cdot\partial)$ are invariant under this rescaling and allow nonlocal operators compatible with VSR while breaking Lorentz symmetry.

Such operators allow mass terms for gauge fields without breaking gauge invariance. These constructions have been studied for non-Abelian fields \cite{Alfaro:2013uva}, electroweak theory \cite{PhysRevD.91.105007}, photons \cite{Cheon:2009zx,PhysRevD.100.055029}, Kalb-Ramond fields \cite{Maluf:2023gbf} and Maxwell-Chern-Simons theory \cite{Belchior:2024eyz,Belchior:2026feh}. Their extension to the graviton gives Very Special Linear Gravity, a theory with two physical graviton polarizations and dispersion relation $\omega^2=|\boldsymbol k|^2+m_g^2$ \cite{Alfaro:2022vpt,Santoni:2023uko,Bonilla:2025mrt}. The preferred direction produces an anisotropic tidal response, and the construction recovers linearized general relativity as $m_g\to0$.

A massive graviton with only two polarizations raises the question of which distinct signatures it could produce in GW detector arms and whether these effects could be observed. We address this question through the detector's response to a quantized gravitational field. The usual in-out formulation computes transition amplitudes between specified asymptotic states. Here, instead, we are interested in the dynamics of the detector for a given initial state of the gravitational field, without selecting a final state. There are different equivalent ways to work in such cases. One, which has received increasing attention in some fields, is the Schwinger-Keldysh approach, which provides a real-time description \cite{Schwinger:1960qe,Keldysh:1964ud}. For a general treatment we refer the reader to \cite{kamenev_2011,Calzetta:2008iqa,Rammer:2007zz}. A second, closely related approach, which we use in this work, is the Feynman-Vernon influence functional \cite{FeynmanVernon}. We closely follow the treatment in Vernon's thesis \cite{Vernon:1959vsj} to integrate out the gravitational field and obtain an effective action for the detector. The resulting dynamics of the detector include dissipative and a noise term induced by the gravitational field. Our aim is to characterize the corrections to both due to the graviton mass $m_g$ and their possible observational signatures.

Therefore, the outline of this work is as follows: Section \ref{sec:action} introduces the graviton and detector-arm actions in VSR. Section \ref{sec:SK} presents the effective action for the gravitational detector using the Feynman-Vernon method. Section \ref{sec:eoms} gives the detector equations of motion and noise correlations for vacuum, thermal and squeezed initial states. Section \ref{sec:conclusions} summarizes the results and their observational scale with some comments. The longer calculations and intermediate steps are collected in four appendices. We use $c=\hbar=1$, $\eta_{\mu\nu}=\operatorname{diag}(-1,+1,+1,+1)$, $k^\mu=(\omega,\boldsymbol k)$, $k^2=-m_g^2$, and $n^\mu=(1,\boldsymbol n)$ with $|\boldsymbol n|=1$.

\section{Action Calculation}
\label{sec:action}

\subsection{VSR action for linearised gravity}

The most general quadratic $SIM(2)$-invariant action for a graviton $h_{\mu \nu}$, which reproduces the Fierz-Pauli form in the massless limit was found in \cite{Alfaro:2022vpt} and it reads
\begin{equation}
S_h
=
\frac{1}{64 \pi G}
\int \mathrm{d}^4 x \;
h_{\mu\nu}(x)\;\mathcal{O}^{\mu\nu\rho\sigma}\;
h_{\rho\sigma}(x)\,,
\label{eq:vslg_action1}
\end{equation}
where the operator $\mathcal O $ is given by

\begin{align}
\mathcal{O}_{\mu\nu\alpha\beta} =&\, 
\partial_{\mu} \partial_{\nu} \eta_{\alpha \beta} + \partial_{\alpha}
\partial_{\beta} \eta_{\mu \nu} - \frac{1}{2} (\partial_{\mu} \partial_{\alpha} \eta_{\nu \beta} + 
\partial_{\mu} \partial_{\beta} \eta_{\nu \alpha}+\partial_{\nu} \partial_{\beta}
\eta_{\mu \alpha} + \partial_{\nu} \partial_{\alpha} \eta_{\mu
\beta}) - \eta_{\mu \nu} \eta_{\alpha \beta} \partial^2 + \frac{1}{2} ( \eta_{\mu \alpha} \eta_{\nu \beta} + 
\eta_{\mu \beta} \eta_{\nu \alpha}) \partial^2 \nonumber \\
&  {-} m^2_g N_{\mu} N_{\nu} \partial_{\alpha}
\partial_{\beta}  {+} \frac{m^2_g}{2} N_{\mu} N_{\alpha} \partial_{\nu}
\partial_{\beta}  {+}  \frac{m^2_g}{2} N_{\mu} N_{\beta} \partial_{\nu}
\partial_{\alpha}  {+} \frac{m^2_g}{2} N_{\nu} N_{\alpha} \partial_{\mu}
\partial_{\beta}  {+} \frac{m^2_g}{2} N_{\nu} N_{\beta} \partial_{\mu}
\partial_{\alpha}  {-}  m^2_g N_{\alpha} N_{\beta} \partial_{\mu} \partial_{\nu} \nonumber \\
&  {+} (m^2_g N_{\mu} N_{\nu} \eta_{\alpha \beta} - \frac{m^2_g}{2} N_{\mu} N_{\alpha} \eta_{\nu \beta} - \frac{m^2_g}{2} N_{\mu} N_{\beta} \eta_{\nu \alpha} - \frac{m^2_g}{2} N_{\nu}
N_{\beta} \eta_{\mu \alpha} - \frac{m^2_g}{2} \eta_{\mu \beta}
N_{\nu} N_{\alpha} + m^2_g N_{\alpha} N_{\beta} \eta_{\mu \nu}) \partial^2 \nonumber \\
&  {-} m^2_g \eta_{\mu \nu} (N_{\alpha} \partial_{\beta}+\partial_{\alpha} N_{\beta})  {+} \frac{m^2_g}{2} \eta_{\mu \alpha} (N_{\nu}
\partial_{\beta} + \partial_{\nu} N_{\beta})  {+} \frac{m^2_g}{2}
\eta_{\mu \beta} (N_{\nu} \partial_{\alpha}+\partial_{\nu} N_{\alpha}) \\
&  {+} \frac{m^2_g}{2} \eta_{\nu \alpha}
(N_{\mu} \partial_{\beta}+\partial_{\mu}
N_{\beta})  {+} \frac{m^2_g}{2} \eta_{\nu \beta} (N_{\mu} \partial_{\alpha}+\partial_{\mu} N_{\alpha})  {-} m^2_g
\eta_{\alpha \beta} (N_{\mu} \partial_{\nu}+\partial_{\mu} N_{\nu}) \nonumber \\
& + m^2_g \eta_{\mu \nu} \eta_{\alpha \beta}
 {-} \frac{m^2_g}{2} \eta_{\mu \alpha} \eta_{\nu \beta}  {-} \frac{m^2_g}{2}
\eta_{\mu \beta} \eta_{\nu \alpha} \nonumber \,,
\label{eq:operator_explicit}
\end{align}
and here $m_g$ is the graviton mass in VSR. Also, we defined
\begin{equation}
    N^\mu \equiv \frac{n^\mu }{n \cdot \partial} \,.
\end{equation}
Notice that when the VSR parameter $m_g$ goes to zero, we recover the standard action for linear gravity.\\
In order to single out the physical degrees of freedom from this action we choose the natural gauge in VSR 
\begin{equation}
    \partial^\mu h_{\mu\nu } = n^\mu h_{\mu\nu} = 0 \,, \qquad h = {h^\mu}_\mu = 0 \,,
\end{equation}
With this choice, the action \eqref{eq:vslg_action1} reduces to
\begin{equation} \label{eq:vslg_action2}
    S_h = \frac{1}{64 \pi G} \int d^4 x [h_{\mu \nu} \partial^2 h^{\mu \nu} - m^2_g h_{\mu \nu} h^{\mu \nu}] \,,
\end{equation}
which gives the correct Klein-Gordon equation in the mostly-plus convention, i.e.,
\begin{equation}
    (\partial^2-m^2_g)h_{\mu\nu} = 0 \,, \, \text{with } \, \partial^2 = -\partial_0^2 + \vec \nabla^2.
\end{equation}
Here, we emphasize that, although we have chosen a specific gauge to simplify the algebraic manipulation of the action, the physical observables and the underlying theory remain strictly gauge invariant. \\
Using integration by parts, we can write Eq.~\eqref{eq:vslg_action2} as
\begin{equation} \label{eq:vslg_action3}
    S_h = \frac{1}{64 \pi G} \int d^4 x [\dot{h}_{\mu \nu} \dot{h}^{\mu \nu} - \vec \nabla h_{\mu \nu} \cdot \vec \nabla h^{\mu \nu} - \,m^2_g h_{\mu \nu} h^{\mu \nu}] \,,
\end{equation}
which corresponds to the action for the graviton to be used.

\subsection{Detector action}

In addition, we consider the action for a gravitational wave detector. We start from the equation for the geodesic distance $\vec\xi$ among two test particles of mass $m_0$. Choosing a frame comoving with one of the two particles, we have
\begin{equation}
\partial_0^2 \xi^i = - R_{i 00 j} \xi^j \,.
\end{equation}
This equation can be obtained from the action
\begin{equation}
S_{\det} = \frac{1}{2} m_0 \int d t \, (\dot{\xi}^i \dot{\xi}^i - \xi^i R_{i 00 j} \xi^j) \,,
\end{equation}
where we have directly included the standard kinetic contribution proportional to the mass of the particle. Using the expression for the linearised Riemann tensor
\begin{equation}
R_{i 00 j} = - \frac{1}{2} (h_{00, i j} + h_{i j, 00} - h_{0 i, 0 j} - h_{0j, 0 i}) \,,
\end{equation}
we can write the action for the detector as
\begin{eqnarray} 
\label{Sdet}
S_{\det} &=& \frac{1}{2} m_0 \int d t \left( \delta_{i j} \dot{\xi}^i
\dot{\xi}^j + \frac{1}{2} (h_{00, i j} + h_{i j, 00} - h_{0 i, 0 j} - h_{0 j,0 i}) \xi^i \xi^j \right)
\end{eqnarray}

\subsection{Full action}

Let us write $h_{\mu \nu}$ in a Fourier space in a covariant form
\begin{equation}
h_{\mu \nu} = \frac{1}{\sqrt{G}} \underset{s}{\sum} \int \frac{d^4 k}{(2
\pi)^4} c_s (k) \varepsilon_{\mu \nu}^{(s)} (k) e^{i k x}
\end{equation}
where $s$ runs over the two possible graviton VSR polarizations and the polarization tensor $\epsilon_{\mu\nu}^s$ satisfies the normalization
\begin{equation}
\epsilon_{\mu \nu}^s (\vec{k}) \epsilon_{s'}^{\mu \nu} (\vec{k}) = {\delta^s}_{s'}
\end{equation}
We can choose a real basis for the polarization tensor to write the action in a compact way. This is just a matter of convenience. In the next steps we will work with gauge invariant quantities as the sum of polarizations, so they can be computed in any basis, and here we choose a convenient one just to express the action in a convenient way. Together with this choice, we will consider that
\begin{equation}
\varepsilon_{\mu \nu}^{(s)} (- k) = \varepsilon_{\mu \nu}^{(s)} (k)
\end{equation}
With this condition, and the reality of $h_{\mu \nu}$ we have that
\begin{equation}
\label{csdef}
c^{\ast}_s (- k) = c_s (k)
\end{equation}
Using all these expressions, the action becomes
\begin{eqnarray}
\label{pasoS}
 S &=& \frac{1}{64 \pi G^2}  \underset{s}{\sum} \int \frac{d^4 k}{(2 \pi)^4} |
   c_s (k) |^2 ((k^0)^2 - \omega^2) \nonumber\\
   & & + \frac{1}{2} m_0 \int
   d t \left( \delta_{i j} \dot{\xi}^i \dot{\xi}^j - \frac{1}{2 \sqrt{G}}
   \underset{s}{\sum} \int \frac{d^4 k}{(2 \pi)^4} e^{i k x} c_s (k)
   \Lambda^{(s)}_{i j} \xi^i \xi^j \right)
\end{eqnarray}
where we defined:
\begin{equation}
\omega^2=| \mathbf{k} |^2 + m^2_g
\end{equation}
and the object
\begin{equation}
\label{lam}
\Lambda^{(s)}_{i j} = \varepsilon_{00}^{(s)} k_i k_j + \varepsilon_{i
j}^{(s)} (k^0)^2 + (\varepsilon_{0 i}^{(s)} k_j + \varepsilon_{0 j}^{(s)} k_i)
k^0
\end{equation}

Notice that the interaction term depends on the polarization only through $\Lambda^{(s)}_{i j}$, which is invariant under the gauge transformation
\begin{equation}
\delta \varepsilon_{\mu \nu}^{(s)} = k_{\mu} \Lambda_{\nu}^{(s)} + k_{\nu}
\Lambda_{\mu}^{(s)}
\end{equation}
for arbitrary $\Lambda_\mu^{(s)}$.

We can see that in the interaction, if we split the $k^0$ integral into the regions
$k^0>0$ and $k^0<0$ and changing the variables $k\to -k$ in the region $k^0<0$, we will have that after using \eqref{csdef}, the fact that $\Lambda^{(s)}_{ij}$ is real due to the reality of the polarization tensor and that it satisfies $\Lambda^{(s)}_{ij}(-k)=\Lambda^{(s)}_{ij}(k)$ we will have
\begin{equation}
\int \frac{d^4 k}{(2\pi)^4}\, e^{ikx}\, c_s(k)\,\Lambda^{(s)}_{ij}
= 2\,\mathrm{Re}\int \frac{d^3 k}{(2\pi)^3}\int_0^{\infty}\frac{dk^0}{2\pi}\,
e^{ikx}\, c_s(k)\,\Lambda^{(s)}_{ij}(k).
\end{equation}
Similarly we can do it for the first term in \eqref{pasoS}, and then we can define
\begin{equation}
q_{\mathbf{k}, s} (t) = \int_0^{\infty} \frac{d k^0}{\sqrt{2}\,\pi} c_s (k^0, \mathbf{k}) e^{- i k^0 t}
\end{equation}
Then, using the Parseval theorem we can write
\begin{equation}
\int_{-\infty}^{\infty} \frac{d k^0}{2\pi} | c_s |^2 = \int d t | q_{\mathbf{k}, s} |^2, \quad \int_{-\infty}^{\infty} \frac{d k^0}{2\pi} (k^0)^2 | c_s |^2 = \int d t | \dot{q}_{\mathbf{k}, s} |^2
\end{equation}
Moreover, to first order, the free action imposes $(k^0)^2=\omega^2$, so in the region $k^0>0$ we have $k^0=\omega$ and
\begin{equation}
\Lambda^{(s)}_{i j} = \varepsilon_{00}^{(s)} k_i k_j + \varepsilon_{i
j}^{(s)} \omega^2 (k) + (\varepsilon_{0 i}^{(s)} k_j + \varepsilon_{0 j}^{(s)}
k_i) \, \omega (k)
\end{equation}

Also, in the following, we will consider the length-scale of the detector to be smaller than the characteristic scale $\lambda$ over which $h_{\mu \nu}$ changes. That allows us to take approximately $e^{i \mathbf{k} \cdot \mathbf{x}}  \approx 1$. With those considerations, the action becomes
\begin{eqnarray}
S &=& \frac{1}{64 \pi G^2}  \underset{s}{\sum} \int d t \int \frac{d^3 k}{(2
\pi)^3} \bigg(| \dot{q}_{\mathbf{k}, s} |^2 - (| \mathbf{k} |^2 + m^2_g) | q_{\mathbf{k}, s} |^2 \bigg) \nonumber\\
& & + \frac{1}{2}
m_0 \int d t \left( \delta_{i j} \dot{\xi}^i \dot{\xi}^j - \frac{1}{
\sqrt{2G}} \underset{s}{\sum} \int \frac{d^3 k}{(2 \pi)^3} \Lambda^{(s)}_{i j}
\Re(q_{\mathbf{k}, s}) \xi^i \xi^j \right)
\end{eqnarray}

Since the interaction only couples to the real part of $q_{\mathbf{k}, s}$, from now on, we will only work with the real part of $q$. Thus, for only one mode $\mathbf{k}$ we will have that the action can be written as
\begin{equation}
S = S_{\xi} + S_q
\end{equation}
with
\begin{eqnarray}
    S_{\xi} &=& \frac{1}{2} m_0 \int^T_{\tau} d t \dot{\xi}^i \dot{\xi}^i \\
    S_q &=& \int^T_{\tau} d t \left[ \frac{1}{2} m_{eff} [\dot{q}^2 -
   \omega^2 q^2] + J q \right]
\end{eqnarray}
where we defined
\begin{eqnarray}
g&=&\frac{1}{\sqrt{2G}}\\
m_{eff} &=& \frac{1}{16 \pi G^2} \\
\omega^2 &=& | \mathbf{k} |^2 +
m^2_g \\
J &=& - \frac{1}{2} m_0 g \Lambda^{(s)}_{i j} (\omega) \xi^i \xi^j
\end{eqnarray}
and we have set explicitly limits of integration for the time integral in the action.

\section{Effective action for the detector}
\label{sec:SK}

In order to construct an effective action for the detector, we will proceed with a path-integral approach. In \cite{PhysRevD.91.105007} was argued that path integral in VSR produces correct expressions of the propagator,
hamiltonian and charge operator in terms of creation and annihilation operators within the canonical
second quantization. Moreover, this approach is simpler and covariant, in contrast with canonical quantization, where the possible existence of second class constraints make this task complicated. Following the Vernon method \cite{Vernon:1959vsj}, we can construct an effective action for $\xi$, which is given by
\begin{equation}
\label{defeffact}
    S_{eff} [\xi] = S_{\xi} [\xi] - S_{\xi} [\xi'] + S_{IF} [\xi,
   \xi']
\end{equation}

where the part
\begin{equation}
\label{defSIF}
    S_{IF} [\xi, \xi'] = - i \ln F [\xi, \xi']
\end{equation}
is defined through the influence functional $F [\xi, \xi']$ which is constructed using the field $q$ as 
\begin{eqnarray}
\label{influencefunc}
F [\xi, \xi'] = \int d q_T d q_{\tau} d q'_T d q'_{\tau} \delta (q_T - q'_T)
\mathcal{D}q\mathcal{D}q' \rho_q (q_{\tau}, q'_{\tau}) \exp [i (S_q [q, \xi] -
S_q [q', \xi'])]    
\end{eqnarray}
and here we will consider a quadratic initial density matrix:
\begin{equation}
\label{generalrho}
\rho_q (q_{\tau}, q'_{\tau}) =\mathcal{N} \exp \left[ - \frac{m_{eff}
\omega}{2} (A (q_{\tau} - a)^2 + A^{\ast} (q_{\tau}' - a^{\ast})^2 - B
(q_{\tau} - a) (q_{\tau}' - a^{\ast})) \right]
\end{equation}
where we choose the coefficients to parametrize in one single expression three possible choices of initial conditions: Vacuum, Thermal and Squeezed states. With this in mind, we choose the coefficients to be
\begin{eqnarray}
    \mathcal{N} &=& \sqrt{\frac{m_{eff} \omega \tanh \left( \frac{\beta
   \omega}{2} \right)}{\pi (\cosh (2 r) - \cos \phi \sinh (2 r))}} \\
   A &=& \frac{\cosh r + e^{i \phi} \sinh r}{\cosh r - e^{i \phi} \sinh r} +
   \frac{\coth (\beta \omega) - 1}{\cosh (2 r) - \cos \phi \sinh (2 r)} \\
   B &=& \frac{2 \csch (\beta \omega)}{\cosh (2 r) - \cos \phi \sinh (2
   r)}
\end{eqnarray}

We observe that in the following limits we recover those states:
\begin{itemize}
    \item Vacuum ($\beta \rightarrow \infty$, and $r \rightarrow 0$, $a \rightarrow 0$)

    Here $\mathcal{N}= \sqrt{\frac{m_{eff} \omega}{\pi}}$, $A = 1$, $B = 0$
    and in consequence
    \begin{equation}
        \rho_q (q_{\tau}, q'_{\tau}) =\mathcal{N} \exp \left[ - \frac{m_{eff}
\omega}{2} \left( q_{\tau}^2 {+ q_{\tau}'}^2 \right) \right]
    \end{equation}

    \item Thermal ($r \rightarrow 0$, $a \rightarrow 0$)

    Here $\mathcal{N}= \sqrt{\frac{m_{eff} \omega \tanh \left( \frac{\beta
\omega}{2} \right)}{\pi}}$, $A = \coth (\beta \omega)$, $B = 2 \csch
(\beta \omega)$ and therefore
\begin{equation}
    \rho_q (q_{\tau}, q'_{\tau}) =\mathcal{N} \exp \left[ - \frac{m_{eff}
\omega}{2} \left( A \left( q_{\tau}^2 {+ q_{\tau}'}^2 \right) - B q_{\tau}
q_{\tau}' \right) \right]
\end{equation}

\item Squeezed ($\beta \rightarrow \infty$)

Here $\mathcal{N}= \sqrt{\frac{m_{eff} \omega}{\pi (\cosh (2 r) - \cos \phi
\sinh (2 r))}}$, $A = \frac{\cosh r + e^{i \phi} \sinh r}{\cosh r - e^{i \phi}
\sinh r}$, $B = 0$ and we have
\begin{equation}
    \rho_q (q_{\tau}, q'_{\tau}) =\mathcal{N} \exp \left[ - \frac{m_{eff}
\omega}{2} (A (q_{\tau} - a)^2 + A^{\ast} (q_{\tau}' - a^{\ast})^2) \right]
\end{equation}
\end{itemize}

with this we can do the integrations, which are reported in appendix \ref{app:effac}. Then, using this influence functional we can employ a typical procedure used in treatments using Schwinger-Keldysh, a change of variable (Keldysh rotation)
\begin{equation}
\xi_{cl} = \frac{1}{2} (\xi + \xi'), \quad \xi_q = \frac{1}{2} (\xi -
\xi')
\end{equation}
to write the effective action for $\xi$. We assume the interaction is switched on at $\tau=0$. It is
\begin{eqnarray}
\label{effectiveactionxi}
S_{eff} [\xi] &=& 2 m_0 \int^T_0 d t \dot{\xi}_{cl}^i
\dot{\xi}_q^i \nonumber\\
& & + \frac{m_0^2 g^2}{(2 \pi)^3 m_{eff}} \int d \Omega
\int^{\infty}_0 d | \mathbf{k} | \frac{| \mathbf{k} |^2}{\omega}
\underset{s}{\sum} \Lambda^{(s)}_{i j} \Lambda^{(s)}_{l m} \int^T_0 d t
\int^T_0 d t' \xi^i_{cl} (t) \xi^j_{cl} (t) \theta (t' - t) \sin
(\omega (t' - t)) \xi^l_{cl} (t') \xi^m_q (t') \nonumber\\
& & + \frac{2 i m_0^2
g^2}{(2 \pi)^3 m_{eff}} \int d \Omega \int^{\infty}_0 d | \mathbf{k}
| \frac{| \mathbf{k} |^2}{\omega} \underset{s}{\sum} \Lambda^{(s)}_{i j}
\Lambda^{(s)}_{l m} \int^T_0 d t \int^T_0 d t' \xi^i_{cl} (t) \xi^j_q
(t) M (t, t') \xi^l_{cl} (t') \xi^m_q (t') \nonumber\\
& & + i \frac{\sqrt{2} m_0 g}{(2
\pi)^3} \int d \Omega \int^{\infty}_0 d | \mathbf{k} | \frac{| \mathbf{k}
|^2 \underset{s}{\sum} \Lambda^{(s)}_{i j}}{(A + A^{\ast} - B)} \int^T_0 d t
\xi^i_{cl} (t) \xi^j_q (t) R (t)
\end{eqnarray}

where we have assumed that the variables $\xi_q$, which represent quantum fluctuations, are much smaller than the classical ones. Thus, we have discarded terms containing more than two of them to construct this effective action. Also, the effective action is described in terms of the following elements:
\begin{eqnarray}
M (t, t') &=& \frac{| A |^2 - \frac{1}{4} B^2 + 1}{2 (2 \Re (A) - B)}
   \cos (\omega (t - t')) - \frac{| A |^2 - \frac{1}{4} B^2 - 1}{2 (2
   \Re (A) - B)} \cos (\omega (t + t')) - \frac{\Im (A)}{2
   \Re (A) - B} \sin (\omega (t + t')) \label{matM}\\
R (t) &=& i \frac{2 (A a + A^{\ast} a^{\ast}) - B (a^{\ast} + a)}{\sqrt{2}}
   \cos (\omega t) - \sqrt{2} \left( | A |^2 - \frac{1}{4} B^2 \right)
   (a^{\ast} - a) \sin (\omega t)
\end{eqnarray}

From now on, for simplicity we will consider the displacements $a$ and $a^\ast$ equal to zero, so $R(t)=0$ and the effective action is simpler. We only need to compute the sum $\underset{s}{\sum} \Lambda^{(s)}_{i j} \Lambda^{(s)}_{l m} $ over polarizations $s$ and its angular integral. We left this computation to the appendix \ref{sec:sumoverlambda}, where for simplicity, we approximate this sum up to order $m_g^2$. Here, we emphasise the fact that this object does not depend on the chosen basis for the polarization tensors. We can compute it in the more convenient basis and the result can be employed here. This fact results in a gauge independent result, which is crucial for the computation to be finite. With this considerations, we get that 
\begin{equation}
\label{sumpolarizations}
\int d \Omega \underset{s}{\sum} \Lambda^{(s)}_{i j} \Lambda^{(s)}_{l m} = 4
\pi \omega^4 \left( \tilde{T}^{(0)}_{ijlm} + \frac{m^2_g}{\omega^2}
\Lambda_{ijlm} + \frac{m^2_g}{\omega^2} \ln \left( \frac{4
\omega^2}{m_g^2} \right) \Xi_{ijlm} \right)
\end{equation}
where we have defined
\begin{equation}
\tilde T^{(0)}_{ijlm}
=\frac15(\delta_{il}\delta_{jm}+\delta_{im}\delta_{jl}
          -\delta_{ij}\delta_{lm})
+\frac1{15}\delta_{ij}\delta_{lm},
\end{equation}
\begin{align}
\Lambda_{ijlm}
={}&\frac{91}{60}
 (\delta_{il}\delta_{jm}+\delta_{im}\delta_{jl}
  -\delta_{ij}\delta_{lm})
+\frac{11}{30}\delta_{ij}\delta_{lm}\nonumber\\
&+\frac{13}{12}
 (\delta_{ij}n_l n_m+\delta_{lm}n_i n_j)\nonumber\\
&-\frac{23}{12}
 (\delta_{il}n_jn_m+\delta_{im}n_jn_l
  +\delta_{jl}n_in_m+\delta_{jm}n_in_l)\nonumber\\
&+\frac{37}{12}n_i n_j n_l n_m,
\end{align}
\begin{align}
\Xi_{ijlm}
={}&-\frac12
 (\delta_{il}\delta_{jm}+\delta_{im}\delta_{jl}
  -\delta_{ij}\delta_{lm})\nonumber\\
&-\frac12(\delta_{ij}n_l n_m+\delta_{lm}n_i n_j)\nonumber\\
&+\frac12
 (\delta_{il}n_jn_m+\delta_{im}n_jn_l
  +\delta_{jl}n_in_m+\delta_{jm}n_in_l)\nonumber\\
&-\frac12 n_i n_j n_l n_m.
\end{align}

Therefore, using \eqref{sumpolarizations} in \eqref{effectiveactionxi}, restoring $m_{eff} = \frac{1}{16 \pi G^2}$ and $g = \frac{1}{\sqrt{2G}}$, passing the integrals in $| \mathbf{k} |$ to $\omega$ and doing algebra that we left to the appendix \ref{sec:effarrange}, the effective action for $\xi$ up to order $m_g^2$ becomes
\begin{eqnarray}
S_{eff} [\xi] &=& - 2 m_0 \int^T_0 d t \frac{\partial^2
\xi_{cl}^i}{\partial t^2} \xi_q^i \nonumber\\
& & - 2 G m_0^2 \int^T_0 d t \frac{\partial^5 (\xi^i_{cl} (t)
\xi^j_{cl} (t))}{\partial t^5} \tilde{T}^{(0)}_{i j l m}
\xi^l_{cl} (t) \xi^m_q (t) \nonumber\\
& & + 2 G m_0^2 m_g^2 \int^T_0 d t \frac{\partial^3 (\xi^i_{cl}
(t) \xi^j_{cl} (t))}{\partial t^3} \tilde{S} _{i j l m}
\xi^l_{cl} (t) \xi^m_q (t) \nonumber\\
& & - 4 G m_0^2 m_g^2 \int^T_0 d t \int^t_0 d t' \frac{\partial^4
\xi^i_{cl} (t') \xi^j_{cl} (t')}{{\partial t'}^4} \ln (m_g | t -
t' |) \Xi_{i j l m} \xi^l_{cl} (t) \xi^m_q (t)
\nonumber\\
& & + i \frac{8 G m_0^2}{\pi} \int^T_0 d t \int^T_0 d t'  \frac{\partial^2
(\xi^i_{cl} (t) \xi^j_q (t))}{\partial t^2} K_{i j l m} (t, t')
\frac{\partial^2 (\xi^l_{cl} (t') \xi^m_q (t'))}{{\partial t'}^2}\label{eact}
\end{eqnarray}

where we defined
\begin{equation}
\tilde{S} _{i j l m} = - \frac{1}{2} \tilde{T}^{(0)}_{i j l m} + \Lambda_{i j
l m} + 2 (\ln (2) - \gamma) \Xi_{i j l m}
\end{equation}
with $\gamma$ the Euler-Mascheroni constant and the kernel $K_{i j l m}$ is defined by
\begin{equation}
K_{i j l m} (t, t') = I_0 (t, t') \tilde{T}^{(0)}_{i j l m} + m_g^2 I_1 (t,
t') \tilde{S} _{i j l m} + 2 m_g^2 I_2 (t, t') \Xi_{i j l m}
\end{equation}
where the objects $I_0$, $I_1$ and $I_2$ are integrals defined by
\begin{eqnarray}
I_0 &=& \int^{\infty}_{m_g} d \omega \omega M (t, t') \\
I_1 &=& \int^{\infty}_{m_g} d \omega \frac{1}{\omega} M (t, t') \\
I_2 &=& \int^{\infty}_{m_g} d \omega \frac{1}{\omega} \left( \gamma + \ln
\left( \frac{\omega}{m_g} \right) \right) M (t, t')
\end{eqnarray}
and $M(t,t')$ is given by \eqref{matM}.

We can arrange the last term of (\ref{eact}) using a Hubbard-Stratonovich transformation in order to define a noise term:
\begin{eqnarray}
\label{finaleffaction}
S_{eff} [\xi] &=& 2 m_0 \int^T_0 d t \xi_q^j \bigg( - \delta_{i j}
\frac{\partial^2}{\partial t^2} - \frac{\partial^2 \mathcal{N}_{i j}}{\partial
t^2} - G m_0  \frac{\partial^5 (\xi^l_{cl}
\xi^m_{cl})}{\partial t^5} \tilde{T}^{(0)}_{l m i j} + G m_0 m_g^2
\frac{\partial^3 (\xi^l_{cl} \xi^m_{cl})}{\partial t^3}
\tilde{S} _{l m i j} \nonumber\\
& & - 2 G m_0 m_g^2 \int^t_0 d t' \frac{\partial^4
(\xi^l_{cl} (t') \xi^m_{cl} (t'))}{{\partial t'}^4} \ln (m_g | t
- t' |) \Xi_{l m i j} \bigg) \xi_{cl}^i \nonumber\\
& & + i \frac{\pi}{8 G} \int^T_0 d t \int^T_0 d t' \mathcal{N}_{i j} (t) K^{-
1}_{i j l m} (t, t') \mathcal{N}_{l m} (t')
\end{eqnarray}

\section{Equations of motion from the effective action and noise term correlation}
\label{sec:eoms}

We can obtain the equation of motion for $\xi_{cl}$ varying respect to $\xi_q$:
\begin{eqnarray}
& & \delta_{i j} \frac{\partial^2 \xi_{cl}^i}{\partial t^2} - \bigg(
\frac{\partial^2 \mathcal{N}_{i j}}{\partial t^2} - G m_0 \tilde{T}^{(0)}_{l m i j} \frac{\partial^5 (\xi^l_{cl} (t)
\xi^m_{cl} (t))}{\partial t^5} + G m_0 m_g^2 \tilde{S} _{l
m i j} \frac{\partial^3 (\xi^l_{cl} (t) \xi^m_{cl}
(t))}{\partial t^3} \nonumber\\
& & - 2 G m_0 m_g^2 \Xi_{l m i j} \int^t_0 d t' \frac{\partial^4
\xi^l_{cl} (t') \xi^m_{cl} (t')}{{\partial t'}^4} \ln (m_g (t -
t')) \bigg) \xi_{cl}^i = 0
\end{eqnarray}
which defines the dynamics of the detector in the presence of the gravitational wave. We observe that in the limit $m_g\to 0$,  the equation simplifies to
\begin{eqnarray}
\delta_{i j} \frac{\partial^2 \xi_{cl}^i}{\partial t^2} - \bigg(
\frac{\partial^2 \mathcal{N}_{i j}}{\partial t^2} - G m_0 \tilde{T}^{(0)}_{l m i j} \frac{\partial^5 (\xi^l_{cl} (t)
\xi^m_{cl} (t))}{\partial t^5} \bigg) \xi_{cl}^i = 0
\end{eqnarray}
We recover the same structure found in \cite{PhysRevD.104.046021}, but with differences in the coefficient of the dissipative term. This difference is due to the fact that we have taken here a sum over polarizations instead of only one.

The noise has a big impact on the equation. Depending on the nature of the initial state they can leave different imprints. The correlation of this variable $\mathcal{N}_{i j }$ is given by
\begin{equation}
\langle \mathcal{N}_{i j} (t) \mathcal{N}_{l m} (t') \rangle = \frac{4
G}{\pi} K_{i j l m} (t, t')
\end{equation}

Since the integrals $I_0$, $I_1$ and $I_2$ are complicated in the general case, we simply compute them for the three cases: vacuum, thermal and squeezed. We leave the explicit computation to the appendix \ref{sec:noiseintegrals}, where we have computed them up to required order to have in $K_{i j l m}$ up to order two, and we have removed constant terms appearing in those objects, since the noise enters in the equation of motion as derivatives of $\mathcal{N}_{i j}$, so they do not contribute. We summarize the values for those cases in the table \ref{tableIs}. Here we remark that for the squeezed states we have extended the analysis in \cite{PhysRevD.104.046021} considering the case when $\phi\neq 0$. For our case it introduces an extra feature.

We notice that in the squeezed state, the term $I_0$ proportional to $\sin{\phi}$ contains a derivative of the delta evaluated in $t+t'$. Since $t$ and $t'$ are such that $t,t'\geq 0$, this term only contributes at the initial time. Thus, in the massless case, the noise during the evolution is not sensitive to the component proportional to $\sin{\phi}$. However, the existence of the mass $m_g$ introduces a contribution proportional to $\sin{\phi}$ in $I_2$, which is multiplied by the tensor $\Xi_{i j l m}$. Therefore, for squeezing parameters independent of the frequency, the sensitivity of the noise to $\sin{\phi}$ is a signature of this model.

\begin{table}[h]
\centering
\large
\renewcommand{\arraystretch}{1.8}
\setlength{\tabcolsep}{6pt}
\resizebox{\textwidth}{!}{%
\begin{tabular}{|c|c|c|c|}
\hline
 & $\displaystyle I_0$ & $\displaystyle I_1$ & $\displaystyle I_2$ \\
\hline

vacuum &
$\begin{gathered}
\\[3pt]
\displaystyle
- \frac{1}{2} PF \left( \frac{1}{(t - t')^2} \right) \\[10pt]
\end{gathered}$
&
$\begin{gathered}
\\[3pt]
\displaystyle
- \frac{1}{2} \ln (m_g | t - t' |) \\[10pt]
\end{gathered}$
&
$\begin{gathered}
\\[3pt]
\displaystyle
\frac{1}{4} (\ln (m_g | t - t' |))^2 \\[10pt]
\end{gathered}$
\\
\hline

thermal &
$\displaystyle
- \frac{\pi^2}{2 \beta^2 \left( \sinh \left( \frac{\pi}{\beta} (t - t')
\right) \right)^2}
$
&
$\displaystyle
- \frac{1}{2} \ln \left( 2 \sinh \left( \frac{\pi}{\beta}
| t - t' | \right) \right)
$
&
$\displaystyle
\begin{gathered}
\\[3pt]
- \frac{\gamma}{2} \ln \left( 2 \sinh \left(
\frac{\pi}{\beta} | t - t' | \right) \right) - \frac{\pi}{2 \beta} | t - t' | (1 - \gamma - \ln (m_g | t - t' |))
\\[3pt]
+ \underset{n = 1}{\overset{\infty}{\sum}} \frac{1}{2 n} \ln \left( \frac{2 \pi
n}{\beta m_g} \right) e^{- \frac{2 \pi n}{\beta} | t - t' |}
\\[5pt]
+ \underset{n = 1}{\overset{\infty}{\sum}} \frac{1}{4 n} \left( e^{\frac{2
\pi n}{\beta} | t - t' |} Ei \left( - \frac{2 \pi n}{\beta} | t - t' |
\right) - e^{- \frac{2 \pi n}{\beta} | t - t' |} Ei \left( \frac{2 \pi
n}{\beta} | t - t' | \right) \right) \\[10pt]
\end{gathered}
$
\\
\hline

squeezed &
$\begin{gathered}
\\[3pt]
\displaystyle
-\frac{1}{2}\cosh(2r)\,
\mathrm{PF}\!\left(\frac{1}{(t-t')^2}\right) \\[5pt]
+\frac{1}{2} \cos \phi \sinh (2 r) PF \left( \frac{1}{(t + t')^2}
\right) \\[5pt]
+ \frac{\pi}{2} \sin \phi \sinh (2 r) \delta' (t + t')
\\[10pt]
\end{gathered}$
&
$\begin{gathered}
\displaystyle
- \frac{1}{2} \cosh (2 r) \ln (m_g | t - t' |) \\[5pt]
+ \frac{1}{2} \cos \phi \sinh (2 r) \ln (m_g | t + t' |)
\end{gathered}$
&
$\displaystyle
\begin{gathered}
\\[3pt]
\frac{1}{4} \cosh (2 r) (\ln (m_g | t - t' |))^2 - \frac{1}{4} \cos \phi \sinh (2 r) (\ln (m_g | t + t' |))^2
\\[5pt]
+ \frac{\pi}{4} \sin \phi \sinh (2 r) \ln (m_g | t + t' |)
\\[5pt]
\end{gathered}
$
\\
\hline

\end{tabular}%
}
\caption{Values of $I_0$, $I_1$ and $I_2$ for vacuum, thermal and squeezed states up to the order required to obtain $K_{i j l m}$ up to order $m_g^2$. Here $PF$ is the Hadamard finite part and $Ei$ the exponential integral function.}
\label{tableIs}
\end{table}

\section{Conclusions}
\label{sec:conclusions}
We have studied the response of a model GW detector to a quantized gravitational field in Very Special Linear Gravity. Working in term of  quantities that are invariant under gauge transformations (see equation \eqref{lam}) is very important for the massive case, since it ensures that the result is free from divergences. 

The influence-functional description relates the detector's causal response and fluctuations to the two physical graviton polarizations. A nonzero mass introduces direction-dependent tensor structures and a logarithmic memory contribution. For the centered Gaussian states considered here, the noise correlations depend on the field state, while the retarded response is the same for vacuum, thermal and squeezed states. The angular corrections vanish smoothly as $m_g\to0$.

The preferred direction suggests looking for orientation-dependent correlations between detector arms, or between detectors sampling the same field. Rotation of the detector relative to this direction could produce a modulation of the correlations. 

The scale of the mass correction can be estimated without specifying the state occupation. Restoring units and writing $E=hf=\hbar\omega$, the characteristic suppression factor for a generic projection with a nonzero logarithmic term is
\begin{equation}
\mathcal R(f)=\left(\frac{m_gc^2}{hf}\right)^2
\ln\!\left[\frac{4(hf)^2}{m_g^2c^4}\right].
\label{eq:conclusion-mass-scale}
\end{equation}
For stationary states, this sets the order of the fractional correction to the noise power, up to geometric coefficients. Taking $m_gc^2=10^{-23}\,\mathrm{eV}$ as an illustrative benchmark gives $\mathcal R\simeq3\times10^{-20}$ at $f=100\,\mathrm{Hz}$ and $\mathcal R\simeq2\times10^{-10}$ at $f=1\,\mathrm{mHz}$. This mass is below the $1.92\times10^{-23}\,\mathrm{eV}/c^2$ propagation bound reported in the GWTC-4.0 parameterized tests \cite{LIGOScientific:2026fcf}; we use that result as context, not as a dedicated bound on VSR. While not being a sensitivity forecast, the larger relative correction in the millihertz band motivates considering LISA \cite{LISA:2024hlh}. The estimate assumes $m_gc^2\ll hf$ and a detector size small compared with the wavelength.

The absolute noise level remains essential. In the vacuum and the freely falling, long-wavelength limit, the equivalent strain power spectrum scales as $S_h^{\rm vac}(f)\sim t_{\rm P}^2f$, where $t_{\rm P}=\sqrt{\hbar G/c^5}$. Its amplitude is of order $10^{-42}\,\mathrm{Hz}^{-1/2}$ at $100\,\mathrm{Hz}$, up to geometric and spectral-convention factors, far below present interferometer sensitivity \cite{PhysRevD.104.046021}. Thermal occupation or squeezing can enhance selected correlations. Fully assessing detectability requires a model of the gravitational state, the interferometer response over the measurement bandwidth, and a comparison with instrumental noise and astrophysical signals. The present results provide kernels for such a calculation; assessing observability requires combining them with a physical source state, the detector transfer function and a finite measurement bandwidth. That analysis however is outside the scope of this paper.

As a final remark, in our work we have kept a general squeezing phase $\phi$, extending also the work done in \cite{PhysRevD.104.046021}, where the phase was set to zero for simplicity. For squeezing parameters independent of the frequency, we found that the graviton mass induces a contribution proportional to $\sin\phi$, with tensor structure $\Xi_{ijlm}$, which persists during the evolution. This makes the noise sensitive to the sign of the squeezing phase, providing a distinctive signature of the model. Squeezed graviton states are created  during inflation \cite{kanno1,kanno2,Mukhanov}. A mechanism to produce squeezed gravitons in a stellar process has been discussed by \cite{Dorlis:2025amf,Dorlis:2025zzz}. There they couple a rotating Black Hole (Kerr) to a massive axion cloud, which is able to radiate squeezed gravitons. 

The relevance of the squeezing phase has also been pointed out in \cite{Mavromatos:2026nin}, where mode-dependent squeezing phases of graviton states such as those of \cite{Dorlis:2025amf,Dorlis:2025zzz} were shown to leave signatures of the gravitational field. Together with our results, this suggests that the squeezing phase may play an important role in identifying different imprints of gravity in general.

\section*{Acknowledgements}
J.A. acknowledges the partial support of the Institute of 	Physics PUC.

A.S. acknowledges support from the European Union's Horizon Europe research and innovation programme under the Marie Sk{\l}odowska-Curie grant agreement No.~101272532 (MALTA4COSMO).

Funded by the European Union. Views and opinions expressed are however those of the author(s) only and do not necessarily reflect those of the European Union or the European Research Executive Agency (REA). Neither the European Union nor the granting authority can be held responsible for them.

Alex Soto acknowledges support from the European Union's Horizon Europe research and innovation programme under the Marie Sk{\l}odowska-Curie grant agreement No 101209941 (Project: BosLong). Also, Alex Soto thanks to Nick Mavromatos for information about graviton squeezed states.

\appendix

\section{Effective action for $\xi$ from the influence functional}
\label{app:effac}

Taking the influence functional $F$ from \eqref{influencefunc} with the initial density matrix defined in \eqref{generalrho} we have that explicitly can be written as
\begin{eqnarray}
F [\xi, \xi'] &=& \int d q_T d q_{\tau} d q'_T d q'_{\tau} \delta (q_T -
   q'_T) \mathcal{D}q\mathcal{D}q' \mathcal{N} \exp \bigg[ i \int^T_{\tau} d t
   \left( \frac{1}{2} m_{eff} \left( \dot{q}^2 - {\dot{q'}}^2 \right) -
   \frac{1}{2} m_{eff} \omega^2 \left( q^2 {- q'}^2 \right) + \left( J
   q - J' q' \right) \right) \nonumber\\
   & & - \frac{m_{eff} \omega}{2} (A (q_{\tau} -
   a)^2 + A^{\ast} (q_{\tau}' - a^{\ast})^2 - B (q_{\tau} - a) (q_{\tau}' -
   a^{\ast})) \bigg]
\end{eqnarray}

Then, we do the change of variables:
\begin{eqnarray}
    x = \frac{1}{\sqrt{2}} (q + q') &,& x' = \frac{1}{\sqrt{2}} (q - q'), \nonumber\\
    K =
\frac{1}{\sqrt{2}} (J + J') &,& K' = \frac{1}{\sqrt{2}} (J - J')
\end{eqnarray}
to write the functional as
\begin{eqnarray}
F [\xi, \xi'] &=& \int d x_T d x_{\tau} d x'_T d x'_{\tau} \delta \left(
\sqrt{2} x_T' \right) \mathcal{D}x\mathcal{D}x' \mathcal{N} \exp \bigg[ i
\int^T_{\tau} d t (m_{eff} \dot{x} \dot{x}' - m_{eff} \omega^2 x
x' + K x' + K' x) \nonumber\\
& & - \frac{m_{eff} \omega}{4} \left( A \left( x_{\tau} +
x_{\tau}' - \sqrt{2} a \right)^2 + A^{\ast} \left( x_{\tau} - x_{\tau}' -
\sqrt{2} a^{\ast} \right)^2 - B \left( x_{\tau} + x_{\tau}' - \sqrt{2} a
\right) \left( x_{\tau} - x_{\tau}' - \sqrt{2} a^{\ast} \right) \right) \bigg] \nonumber\\
\end{eqnarray}

Integrating by parts the first term in the exponential and integrating over $x(t)$ we get
\begin{eqnarray}
F [\xi, \xi'] &=& \int d x_T d x_{\tau} d x'_T d x'_{\tau} \delta \left(
\sqrt{2} x_T' \right) \mathcal{D}x' \delta (m_{eff} (\ddot{x}' +
\omega^2 x') - K') \mathcal{N} \exp \bigg[ i \int^T_{\tau} d t K x' + i
m_{eff} (x_T \dot{x}_T' - x_{\tau} \dot{x}_{\tau}') \nonumber\\
& & - \frac{m_{eff} \omega}{4} \left( A \left( x_{\tau} + x_{\tau}' -
\sqrt{2} a \right)^2 + A^{\ast} \left( x_{\tau} - x_{\tau}' - \sqrt{2}
a^{\ast} \right)^2 - B \left( x_{\tau} + x_{\tau}' - \sqrt{2} a \right) \left(
x_{\tau} - x_{\tau}' - \sqrt{2} a^{\ast} \right) \right) \bigg] \nonumber\\
\end{eqnarray}
The delta that appeared from this integration defines  $x'(t)$ to be the solution of the classical equation 
\begin{equation}
\label{classicx}
    m_{eff} (\ddot{x}' +
\omega^2 x') - K' = 0
\end{equation} 
Then, we do the integral in $x'(t)$, so it fixes $x'$ to be the solution of \eqref{classicx} and then we do the integrals over $x_T$ and $x_\tau$, and after arranging we get
\begin{eqnarray}
\label{pasoFx}
F [\xi, \xi'] &=& \int d x'_T d x'_{\tau} \delta \left( \sqrt{2} x_T' \right)
\delta (m_{eff} \dot{x}_T') \mathcal{N} \sqrt{\frac{4
\pi}{m_{eff} \omega (A + A^{\ast} - B)}} \exp \left[ -
\frac{m_{eff} \omega \left( | A |^2 - \frac{1}{4} B^2 \right)}{2 (A +
A^{\ast} - B)} (a^{\ast} - a)^2 \right] \nonumber\\
& & \times \exp \bigg[ i \int^T_{\tau} d t K
x' - \frac{m_{eff} \omega \left( | A |^2 - \frac{1}{4} B^2
\right)}{(A + A^{\ast} - B)} {x_{\tau}'}^2 - i \frac{m_{eff} (A^{\ast}
- A)}{(A + A^{\ast} - B)} x_{\tau}' \dot{x}_{\tau}' - \frac{m_{eff}}{\omega (A + A^{\ast} - B)} {\dot{x}_{\tau'}}^2 \nonumber\\
& & - \frac{m_{eff} \omega \left( | A |^2 - \frac{1}{4} B^2 \right)}{(A +
A^{\ast} - B)} \sqrt{2} (a^{\ast} - a) x_{\tau}' - i \frac{m_{eff} (2
(A a + A^{\ast} a^{\ast}) - B (a^{\ast} + a))}{\sqrt{2} (A + A^{\ast} - B)}
\dot{x}_{\tau}' \bigg]
\end{eqnarray}

With this expression, from the deltas, we observe that the solution for $x'(t)$ must satisfy that evaluated in $T$ is zero. It means that the solution $x'$ which solves the classical equation \eqref{classicx} and satisfies the conditions in $T$ is
\begin{equation}
x' (t) = \frac{1}{m_{eff} \omega} \int^T_t \sin (\omega (t' -
t)) K' (t') d t'
\end{equation}
Inserting this solution in \eqref{pasoFx} and rewriting
\begin{equation}
\int^T_t \sin (\omega (t' - t)) K' (t') d t' = \int^T_{\tau} \theta (t' - t)
\sin (\omega (t' - t)) K' (t') d t'
\end{equation}
we have that the functional $F$ can be written as
\begin{eqnarray}
F [\xi, \xi'] &=& \int d x'_T d x'_{\tau} \delta \left( \sqrt{2} x_T' \right)
\delta (m_{eff} \dot{x}_T') \mathcal{N} \sqrt{\frac{4
\pi}{m_{eff} \omega (A + A^{\ast} - B)}} \exp \left[ -
\frac{m_{eff} \omega \left( | A |^2 - \frac{1}{4} B^2 \right)}{2 (A +
A^{\ast} - B)} (a^{\ast} - a)^2 \right] \nonumber\\
& & \times \exp \bigg[ \frac{i}{m_{eff}
\omega} \int^T_{\tau} d t \int^T_{\tau} d t' K (t) \theta (t' - t) \sin
(\omega (t' - t)) K' (t') \nonumber\\
& & - \frac{1}{m_{eff} \omega} \int^T_{\tau} d t
\int^T_{\tau} d t' K' (t) M (t, t') K' (t') + \frac{1}{A + A^{\ast} - B}
\int^T_{\tau} R (t) K' (t) d t \bigg]
\end{eqnarray}
where we defined
\begin{eqnarray}
M (t, t') &=& \frac{| A |^2 - \frac{1}{4} B^2 + 1}{2 (2 \Re (A) - B)}
\cos (\omega (t - t')) - \frac{| A |^2 - \frac{1}{4} B^2 - 1}{2 (2 \Re
(A) - B)} \cos (\omega (t + t' - 2 \tau)) - \frac{\Im (A)}{(2 \Re
(A) - B)} \sin (\omega (t - t')) \nonumber\\
& & - \frac{\Im (A)}{(2 \Re (A) - B)}
\sin (\omega (t + t' - 2 \tau)) \\
R (t) &=& i \frac{2 (A a + A^{\ast} a^{\ast}) - B (a^{\ast} + a)}{\sqrt{2}}
\cos (\omega (t - \tau)) - \sqrt{2} \left( | A |^2 - \frac{1}{4} B^2 \right)
(a^{\ast} - a) \sin (\omega (t - \tau))
\end{eqnarray}

Considering that the integrals in $d x'_T d x'_{\tau}$ with the deltas $\delta
\left( \sqrt{2} x_T' \right) \delta (m_{eff} \dot{x}_T')$ just fix the
initial conditions for $x'$, and also change the normalization constant $\mathcal{N}$ due to the jacobian originated in the change of variable from $d x'_{\tau}$ to $d \dot{x}_T'$, we can simply remove the integrals and deltas. Then, we
take $\tau\to 0$ and with that we notice that in $\int^T_0 d t \int^T_0 d t' K' (t) M (t, t') K' (t')$ the term with $\sin
(\omega (t - t'))$ vanishes since exchanging $t \leftrightarrow t'$ gives the
same integral with a minus sign since $t$ and $t'$ are dummy integration variables. Therefore, we can write
\begin{eqnarray}
\label{Falmostfinal}
F [\xi, \xi'] &=& \mathcal{N} \sqrt{\frac{4 \pi}{m_{eff} \omega (A +
A^{\ast} - B)}} \exp \left[ - \frac{m_{eff} \omega \left( | A |^2 -
\frac{1}{4} B^2 \right)}{2 (A + A^{\ast} - B)} (a^{\ast} - a)^2 \right] \nonumber\\
& & \times \exp
\bigg[ \frac{i}{m_{eff} \omega} \int^T_0 d t \int^T_0 d t' K (t) \theta
(t' - t) \sin (\omega (t' - t)) K' (t') \nonumber\\
& & - \frac{1}{m_{eff} \omega}
\int^T_0 d t \int^T_0 d t' K' (t) M (t, t') K' (t') + \frac{1}{A + A^{\ast} -
B} \int^T_0 R (t) K' (t) d t \bigg]
\end{eqnarray}
with
\begin{eqnarray}
M (t, t') &=& \frac{| A |^2 - \frac{1}{4} B^2 + 1}{2 (2 \Re (A) - B)}
   \cos (\omega (t - t')) - \frac{| A |^2 - \frac{1}{4} B^2 - 1}{2 (2
   \Re (A) - B)} \cos (\omega (t + t')) - \frac{\Im (A)}{2
   \Re (A) - B} \sin (\omega (t + t')) \\
R (t) &=& i \frac{2 (A a + A^{\ast} a^{\ast}) - B (a^{\ast} + a)}{\sqrt{2}}
   \cos (\omega t) - \sqrt{2} \left( | A |^2 - \frac{1}{4} B^2 \right)
   (a^{\ast} - a) \sin (\omega t)
\end{eqnarray}

Taking the logarithm of this expression for all the modes, and from the equations \eqref{defeffact} and \eqref{defSIF} we can construct the effective action for $\xi$:
\begin{eqnarray}
S_{eff} [\xi] &=& \frac{1}{2} m_0 \int^T_0 d t \left( \dot{\xi}^i
   \dot{\xi}^i - {\dot{\xi'}}^i {\dot{\xi'}}^i \right) + \underset{s}{\sum}
   \int \frac{d^3 k}{(2 \pi)^3} \frac{1}{m_{eff} \omega} \int^T_0 d t
   \int^T_0 d t' K (t) \theta (t' - t) \sin (\omega (t' - t)) K' (t') \nonumber\\
   & & + \underset{s}{\sum} \int \frac{d^3 k}{(2 \pi)^3} \frac{i}{m_{eff}
   \omega} \int^T_0 d t \int^T_0 d t' K' (t) M (t, t') K' (t') \nonumber\\
   & & - \underset{s}{\sum} \int \frac{d^3 k}{(2 \pi)^3} \frac{i}{2 \Re (A) -
   B} \int^T_0 R (t) K' (t) d t
\end{eqnarray}
where we discarded the first line in \eqref{Falmostfinal} since it is just a constant which is irrelevant to the classical equation of motion of $\xi$.

Now, we return the $K$ and $K'$ to the original variable $J$ and $J'$ that contains $\xi$ explicitly and the effective action reads
\begin{eqnarray}
S_{eff} [\xi] &=& \frac{1}{2} m_0 \int^T_0 d t \left( \dot{\xi}^i
\dot{\xi}^i - {\dot{\xi'}}^i {\dot{\xi'}}^i \right) \nonumber\\
& & + \underset{s}{\sum} \int
\frac{d^3 k}{(2 \pi)^3} \frac{m_0^2 g^2}{8 m_{eff} \omega} \int^T_0 d t
\int^T_0 d t' \Lambda^{(s)}_{i j} (\omega) \Lambda^{(s)}_{l m} (\omega) \left(
\xi^i (t) \xi^j (t) {+ \xi'}^i (t) {\xi'}^j (t) \right) \theta (t' - t) \sin
(\omega (t' - t)) \nonumber\\
& & \times \left( \xi^l (t') \xi^m (t') {- \xi'}^l (t') {\xi'}^m (t')
\right) \nonumber\\
& & + \underset{s}{\sum} \int \frac{d^3 k}{(2 \pi)^3} \frac{i m_0^2 g^2}{8
m_{eff} \omega} \int^T_0 d t \int^T_0 d t' \Lambda^{(s)}_{i j} (\omega)
\Lambda^{(s)}_{l m} (\omega) \left( \xi^i (t) \xi^j (t) {- \xi'}^i (t)
{\xi'}^j (t) \right) M (t, t') \nonumber\\
& & \times \left( \xi^l (t') \xi^m (t') {- \xi'}^l (t')
{\xi'}^m (t') \right) \nonumber\\
& & + \underset{s}{\sum} \int \frac{d^3 k}{(2 \pi)^3}
\frac{i m_0 g}{2 \sqrt{2} (A + A^{\ast} - B)} \int^T_0 d t \Lambda^{(s)}_{i j}
(\omega) \left( \xi^i (t) \xi^j (t) {- \xi'}^i (t) {\xi'}^j (t) \right) R (t)
\end{eqnarray}

Now, we do a change of variables:
\begin{equation}
\xi_{cl} = \frac{1}{2} (\xi + \xi'), \quad \xi_q = \frac{1}{2} (\xi -
\xi')
\end{equation}
and we discard elements with more than two $\xi_q$, under the assumption that those variables, which represent quantum fluctuations are much smaller than the classical ones. Also, we use that $\Lambda_{i j}$
is symmetric under $i \leftrightarrow j$ and with that, the effective action becomes
\begin{eqnarray}
S_{eff} [\xi] &=& 2 m_0 \int^T_0 d t \dot{\xi}_{cl}^i
\dot{\xi}_q^i + \frac{m_0^2 g^2}{m_{eff}} \int \frac{d^3 k}{(2 \pi)^3}
\frac{\underset{s}{\sum} \Lambda^{(s)}_{i j} \Lambda^{(s)}_{l m}}{\omega}
\int^T_0 d t \int^T_0 d t' \xi^i_{cl} (t) \xi^j_{cl} (t) \theta
(t' - t) \sin (\omega (t' - t)) \xi^l_{cl} (t') \xi^m_q (t') \nonumber\\
& & + \frac{2
i m_0^2 g^2}{m_{eff}} \int \frac{d^3 k}{(2 \pi)^3}
\frac{\underset{s}{\sum} \Lambda^{(s)}_{i j} \Lambda^{(s)}_{l m}}{\omega}
\int^T_0 d t \int^T_0 d t' \xi^i_{cl} (t) \xi^j_q (t) M (t, t')
\xi^l_{cl} (t') \xi^m_q (t') \nonumber\\
& & + i \sqrt{2} m_0 g \int \frac{d^3 k}{(2
\pi)^3} \frac{\underset{s}{\sum} \Lambda^{(s)}_{i j}}{(A + A^{\ast} - B)}
\int^T_0 d t \xi^i_{cl} (t) \xi^j_q (t) R (t)
\end{eqnarray}
which is used in the main text writing the momentum integral in spherical coordinates.

\section{Angular integral of $\underset{s}{\sum} \Lambda^{(s)}_{i j} \Lambda^{(s)}_{l m} $}
\label{sec:sumoverlambda}

Define the rectangular kernel
\begin{equation}
\widehat{k}^{\,\mu}{}_i
  =\omega\delta^\mu_i+k_i\delta^\mu_0,
\qquad
\widehat{k}^{\,\mu}{}_i k_\mu=0,
\label{eq:app-kernel}
\end{equation}
where the second identity follows from \(k_0=-\omega\).  The $\Lambda-$tensor can then be written as the double kernel contraction
\begin{equation}
\Lambda^{(s)}_{ij}
  =\widehat{k}^{\,\mu}{}_i\widehat{k}^{\,\nu}{}_j
   \varepsilon^{(s)}_{\mu\nu}.
\label{eq:app-lambda-kernel}
\end{equation}
Define the polarization sum
\begin{equation}
S_{\mu\nu\rho\sigma}
  \equiv\sum_s\varepsilon^{(s)}_{\mu\nu}
                    \varepsilon^{(s)}_{\rho\sigma}.
\label{eq:app-polarization-sum}
\end{equation}
The two physical polarizations are normalized by
\begin{equation}
\varepsilon^{(s)\mu\nu}\varepsilon^{(s')}_{\mu\nu}
  =\delta^{ss'},
\qquad
S^{\mu\nu}{}_{\mu\nu}=2.
\label{eq:app-polarization-normalization}
\end{equation}
Its transversality and tracelessness are
\begin{equation}
k^\mu S_{\mu\nu\rho\sigma}
  =n^\mu S_{\mu\nu\rho\sigma}
  =\eta^{\mu\nu}S_{\mu\nu\rho\sigma}=0.
\label{eq:app-polarization-properties}
\end{equation}
Thus
\begin{equation}
\sum_s\Lambda^{(s)}_{ij}\Lambda^{(s)}_{lm}
 =\widehat{k}^{\,\mu}{}_i\widehat{k}^{\,\nu}{}_j
  \widehat{k}^{\,\rho}{}_l\widehat{k}^{\,\sigma}{}_m
  S_{\mu\nu\rho\sigma}
 =\frac{1}{2}\left(
   \mathcal P_{il}\mathcal P_{jm}
  +\mathcal P_{im}\mathcal P_{jl}
  -\mathcal P_{ij}\mathcal P_{lm}\right),
\label{eq:Lambda-Lambda}
\end{equation}
where
\begin{equation}
\mathcal P_{ij}
 =\omega^2\delta_{ij}
  -\left(1+\frac{m_g^2}{(n\cdot k)^2}\right)k_i k_j
  +\omega\frac{m_g^2}{(n\cdot k)^2}(n_i k_j+n_j k_i)
  -\omega^2\frac{m_g^2}{(n\cdot k)^2}n_i n_j.
\label{eq:P-def}
\end{equation}
With
\begin{equation}
\epsilon=\frac{m_g^2}{\omega^2},
\qquad
L = \ln \left (\frac{\omega +|\vec k|}{\omega - |\vec k|} \right) =\ln\!\left(\frac{4\omega^2}{m_g^2}\right)
  -\frac{m_g^2}{2\omega^2}
  +O\!\left(\frac{m_g^4}{\omega^4}\right),
\label{eq:app-small-mass-parameters}
\end{equation}
the angular integral through first order can be written in the form used in Eq.~\eqref{sumpolarizations}:
\begin{equation}
\int d\Omega\sum_s\Lambda^{(s)}_{ij}\Lambda^{(s)}_{lm}
=4\pi\omega^4\left[
\tilde T^{(0)}_{ijlm}
+\frac{m_g^2}{\omega^2}\Lambda_{ijlm}
+\frac{m_g^2}{\omega^2}\ln\!\left(\frac{4\omega^2}{m_g^2}\right)
 \Xi_{ijlm}
+O\!\left(\frac{m_g^4}{\omega^4}L\right)
\right],
\label{sumlambda}
\end{equation}
where
\begin{equation}
\tilde T^{(0)}_{ijlm}
=\frac15(\delta_{il}\delta_{jm}+\delta_{im}\delta_{jl}
          -\delta_{ij}\delta_{lm})
+\frac1{15}\delta_{ij}\delta_{lm},
\end{equation}
and the nonlogarithmic and logarithmic corrections are, respectively,
\begin{align}
\Lambda_{ijlm}
={}&\frac{91}{60}
 (\delta_{il}\delta_{jm}+\delta_{im}\delta_{jl}
  -\delta_{ij}\delta_{lm})
+\frac{11}{30}\delta_{ij}\delta_{lm}\nonumber\\
&+\frac{13}{12}
 (\delta_{ij}n_l n_m+\delta_{lm}n_i n_j)\nonumber\\
&-\frac{23}{12}
 (\delta_{il}n_jn_m+\delta_{im}n_jn_l
  +\delta_{jl}n_in_m+\delta_{jm}n_in_l)\nonumber\\
&+\frac{37}{12}n_i n_j n_l n_m,
\label{eq:app-Lambda-tensor}
\end{align}
\begin{align}
\Xi_{ijlm}
={}&-\frac12
 (\delta_{il}\delta_{jm}+\delta_{im}\delta_{jl}
  -\delta_{ij}\delta_{lm})\nonumber\\
&-\frac12(\delta_{ij}n_l n_m+\delta_{lm}n_i n_j)\nonumber\\
&+\frac12
 (\delta_{il}n_jn_m+\delta_{im}n_jn_l
  +\delta_{jl}n_in_m+\delta_{jm}n_in_l)\nonumber\\
&-\frac12 n_i n_j n_l n_m.
\label{eq:app-Xi-tensor}
\end{align}
Both tensors depend only on $\delta_{ij}$ and the preferred direction
$n_i$.  Replacing $L$ by $\ln(4\omega^2/m_g^2)$ in the first-order
correction changes the result only at order $\epsilon^2$, within the
remainder retained in Eq.~\eqref{sumlambda}.
The mass-dependent correction vanishes as $\epsilon\ln(1/\epsilon)$ when $\epsilon\to0$ (which corresponds to vanishing $m_g$), so the angular integral approaches its massless value smoothly.

\section{Effective action arrangements}
\label{sec:effarrange}

From equation \eqref{effectiveactionxi}, with $a=a^\ast=0$, when we insert \eqref{sumpolarizations}, we use $m_{eff} = \frac{1}{16 \pi G^2}$ and $g = \frac{1}{\sqrt{2G}}$ and we pass the integral in $| \mathbf{k} |$ to $\omega$ using that $\omega=\sqrt{| \mathbf{k} |^2 + m_g^2}$, we have that the effective action becomes
\begin{eqnarray}
S_{eff} [\xi] &=& 2 m_0 \int^T_0 d t \dot{\xi}_{cl}^i
\dot{\xi}_q^i \nonumber\\
& & + \frac{4 G m_0^2}{\pi} \int^{\infty}_{m_g} \sqrt{\omega^2 - m_g^2} d \omega
\left( \omega^4 \tilde{T}^{(0)}_{i j l m} + m^2_g \omega^2 \left( \Lambda_{i j
l m} + \ln \left( \frac{4 \omega^2}{m_g^2} \right) \Xi_{i j l m} \right)
\right) \nonumber\\
& & \times \int^T_0 d t \int^T_0 d t' \xi^i_{cl} (t) \xi^j_{cl} (t)
\theta (t' - t) \sin (\omega (t' - t)) \xi^l_{cl} (t') \xi^m_q (t') \nonumber\\
& & + i \frac{8 G m_0^2}{\pi} \int^{\infty}_{m_g} \sqrt{\omega^2 - m_g^2} d
\omega \left( \omega^4 \tilde{T}^{(0)}_{i j l m} + m^2_g \omega^2 \left(
\Lambda_{i j l m} + \ln \left( \frac{4 \omega^2}{m_g^2} \right) \Xi_{i j l m}
\right) \right) \nonumber\\
& & \times \int^T_0 d t \int^T_0 d t' \xi^i_{cl} (t) \xi^j_q (t) M
(t, t') \xi^l_{cl} (t') \xi^m_q (t')
\end{eqnarray}
Since the sum of polarizations used was approximated up to order $m_g^2$ we must work at that level of approximation, so expanding the terms up to that order we get
\begin{eqnarray}
S_{eff} [\xi] &=& 2 m_0 \int^T_0 d t \dot{\xi}_{cl}^i
\dot{\xi}_q^i \nonumber\\
& & + \frac{4 G m_0^2}{\pi} \int^T_0 d t \int^T_0 d t' \xi^i_{cl} (t)
\xi^j_{cl} (t) \theta (t' - t) \int^{\infty}_{m_g} d \omega \omega^5
\sin (\omega (t' - t)) \tilde{T}^{(0)}_{i j l m} \xi^l_{cl} (t')
\xi^m_q (t') \nonumber\\
& & + \frac{4 G m_0^2 m_g^2}{\pi} \int^T_0 d t \int^T_0 d t' \xi^i_{cl}
(t) \xi^j_{cl} (t) \theta (t' - t) \int^{\infty}_{m_g} d \omega
\omega^3 \sin (\omega (t' - t)) \tilde{S}' _{i j l m} \xi^l_{cl} (t')
\xi^m_q (t') \nonumber\\
& & + \frac{8 G m_0^2 m_g^2}{\pi} \int^T_0 d t \int^T_0 d t' \xi^i_{cl}
(t) \xi^j_{cl} (t) \theta (t' - t) \int^{\infty}_{m_g} d \omega
\omega^3 \ln \left( \frac{\omega}{m_g} \right) \sin (\omega (t' - t)) \Xi_{i j
l m} \xi^l_{cl} (t') \xi^m_q (t') \nonumber\\
& & + i \frac{8 G m_0^2}{\pi} \int^T_0 d t \int^T_0 d t' \xi^i_{cl} (t)
\xi^j_q (t) \int^{\infty}_{m_g} d \omega \left( \omega^5 \tilde{T}^{(0)}_{i j
l m} + m_g^2 \omega^3 \tilde{S}'_{i j l m} + 2 m_g^2 \omega^3 \ln \left(
\frac{\omega}{m_g} \right) \Xi_{i j l m} \right) \nonumber\\
& &\times M (t, t') \xi^l_{cl}
(t') \xi^m_q (t')
\end{eqnarray}
where we defined
\begin{equation}
\tilde{S}' _{i j l m} = - \frac{1}{2} \tilde{T}^{(0)}_{i j l
m} + \Lambda_{i j l m} + 2 \Xi_{i j l m} \ln (2)
\end{equation}

The integrals can be rearranged extracting time derivatives as:
\begin{eqnarray}
\label{seffdepaso}
S_{eff} [\xi] &=& 2 m_0 \int^T_0 d t \dot{\xi}_{cl}^i
\dot{\xi}_q^i \nonumber\\
& & + \frac{4 G m_0^2}{\pi} \int^T_0 d t \int^T_0 d t' \xi^i_{cl} (t)
\xi^j_{cl} (t) \theta (t' - t) \frac{\partial^5}{\partial t^5}
\int^{\infty}_{m_g} d \omega \cos (\omega (t' - t)) \tilde{T}^{(0)}_{i j l m}
\xi^l_{cl} (t') \xi^m_q (t') \nonumber\\
& & - \frac{4 G m_0^2 m_g^2}{\pi} \int^T_0 d t \int^T_0 d t' \xi^i_{cl}
(t) \xi^j_{cl} (t) \theta (t' - t) \frac{\partial^3}{\partial t^3}
\int^{\infty}_{m_g} d \omega \cos (\omega (t' - t)) \tilde{S}'_{i j l m}
\xi^l_{cl} (t') \xi^m_q (t') \nonumber\\
& & - \frac{8 G m_0^2 m_g^2}{\pi} \int^T_0 d t \int^T_0 d t' \xi^i_{cl}
(t) \xi^j_{cl} (t) \theta (t' - t) \frac{\partial^3}{\partial t^3}
\int^{\infty}_{m_g} d \omega \ln \left( \frac{\omega}{m_g} \right) \cos
(\omega (t' - t)) \Xi_{i j l m} \xi^l_{cl} (t') \xi^m_q (t') \nonumber\\
& & + i \frac{8 G m_0^2}{\pi} \int^T_0 d t \int^T_0 d t' \xi^i_{cl} (t)
\xi^j_q (t) \frac{\partial^2}{\partial t^2} \frac{\partial^2}{{\partial t'}^2}
K_{i j l m} (t, t') \xi^l_{cl} (t') \xi^m_q (t')
\end{eqnarray}
with the definition of
\begin{equation}
K_{i j l m} (t, t') = \int^{\infty}_{m_g} d \omega \left( \omega
\tilde{T}^{(0)}_{i j l m} + m_g^2 \frac{1}{\omega} \tilde{S}'_{i j l m} + 2
m_g^2 \frac{1}{\omega} \ln \left( \frac{\omega}{m_g} \right) \Xi_{i j l m}
\right) M (t, t')
\end{equation}

Let us to see the integrals
\begin{equation}
\int^{\infty}_{m_g} d \omega \cos (\omega (t' - t)), \quad \textrm{and} \quad \int^{\infty}_{m_g}
d \omega \ln \left( \frac{\omega}{m_g} \right) \cos (\omega (t' - t))
\end{equation}
For the first one, splitting in two pieces:
\begin{equation}
\int^{\infty}_{m_g} d \omega \cos (\omega (t' - t)) = \int^{\infty}_0 d
\omega \cos (\omega (t' - t)) - \int^{m_g}_0 d \omega \cos (\omega (t' - t))
\end{equation}
we observe that the second term is finite, and the first one, can be understood in the distributional sense via the relation:
\begin{equation}
\label{relationforkint}
\int_{- \infty}^{\infty} d k k^n e^{i k (t - t')} = 2 \pi \frac{1}{i^n}
\delta^{(n)} (t - t')
\end{equation}
so we have that the derivative of that term, which is what we have in \eqref{seffdepaso}, becomes
\begin{equation}
\frac{\partial^5}{\partial t^5} \int^{\infty}_{m_g} d \omega \cos (\omega (t'
- t)) = \pi \frac{\partial^5}{\partial t^5} \delta (t - t') -
\frac{\partial^5}{\partial t^5} \frac{\sin (m_g (t' - t))}{t' - t}
\end{equation}
If we expand the second term the result is of order higher than $m_g^2$, so we can ignore it.
For the second integral, we proceed similarly, splitting in two pieces and changing variable:
\begin{equation}
\int^{\infty}_{m_g} d \omega \ln \left( \frac{\omega}{m_g} \right) \cos
(\omega (t' - t)) = m_g \int^{\infty}_0 d x \ln (x) \cos (x m_g (t' - t)) - m_g \int^1_0 d x \ln
(x) \cos (x m_g (t' - t))
\end{equation}
As before, the second term is finite, and the first one can be arranged as:
\begin{equation}
\int^{\infty}_0 d x \ln (x) \cos (x m_g (t' - t)) = - \int^{\infty}_0 d x
\frac{1}{m_g} \frac{\partial}{\partial t} \left[ \ln (x) \frac{\sin (x m_g (t'
- t))}{x} \right]
\end{equation}
Interchanging integral and derivative, that integral can be done easily and after deriving we obtain
\begin{equation}
\int^{\infty}_{m_g} d \omega \ln \left( \frac{\omega}{m_g} \right) \cos
(\omega (t' - t)) = - \pi \gamma \delta (t' - t) + \frac{\pi}{2}
\frac{\partial sgn (t' - t) \ln | m_g (t' - t) |}{\partial t} +
\frac{Si (m_g (t' - t))}{t' - t}
\end{equation}
where $\gamma$ is the Euler-Mascheroni constant and $Si$ is the Sine Integral function. Taking the third derivative or the fifth derivative respect to time, we will found that the last term will result in a term of order higher than $m_g^2$, so we can neglect it. With this observation, we found that the finite term will vanish when we take time derivatives. Therefore, with that, we effective action becomes
\begin{eqnarray}
\label{seffpaso2}
S_{eff} [\xi] &=& 2 m_0 \int^T_0 d t \dot{\xi}_{cl}^i
\dot{\xi}_q^i \nonumber\\
& & + 4 G m_0^2 \int^T_0 d t \int^T_0 d t' \xi^i_{cl} (t)
\xi^j_{cl} (t) \theta (t' - t) \frac{\partial^5}{\partial t^5} \delta
(t - t') \tilde{T}^{(0)}_{i j l m} \xi^l_{cl} (t') \xi^m_q (t') \nonumber\\
& & - 4 G m_0^2 m_g^2 \int^T_0 d t \int^T_0 d t' \xi^i_{cl} (t)
\xi^j_{cl} (t) \theta (t' - t) \frac{\partial^3}{\partial t^3} \delta
(t - t') \tilde{S}_{i j l m} \xi^l_{cl} (t') \xi^m_q (t') \nonumber\\
& & - 4 G m_0^2 m_g^2 \int^T_0 d t \int^T_0 d t' \xi^i_{cl} (t)
\xi^j_{cl} (t) \theta (t' - t) \frac{\partial^4 sgn (t' - t) \ln
(m_g | t' - t |)}{\partial t^4} \Xi_{i j l m} \xi^l_{cl} (t') \xi^m_q
(t')
\nonumber\\
& & + i \frac{8 G m_0^2}{\pi} \int^T_0 d t \int^T_0 d t' \xi^i_{cl} (t)
\xi^j_q (t) \frac{\partial^2}{\partial t^2} \frac{\partial^2}{{\partial t'}^2}
K_{i j l m} (t, t') \xi^l_{cl} (t') \xi^m_q (t')
\end{eqnarray}
where we redefine:
\begin{equation}
\tilde{S}_{i j l m} = \tilde{S}'_{i j l m} - 2\gamma \Xi_{i j l m} = - \frac{1}{2} \tilde{T}^{(0)}_{i j l m} + \Lambda_{i j
l m} + 2 (\ln (2) - \gamma) \Xi_{i j l m}
\end{equation}
and with it,
\begin{equation}
K_{i j l m} (t, t') = I_0 (t, t') \tilde{T}^{(0)}_{i j l m} + m_g^2 I_1 (t,
t') \tilde{S} _{i j l m} + 2 m_g^2 I_2 (t, t') \Xi_{i j l m}
\end{equation}
where we define
\begin{eqnarray}
I_0 &=& \int^{\infty}_{m_g} d \omega \omega M (t, t') \\
I_1 &=& \int^{\infty}_{m_g} d \omega \frac{1}{\omega} M (t, t') \\
I_2 &=& \int^{\infty}_{m_g} d \omega \frac{1}{\omega} \left( \gamma + \ln \left(
\frac{\omega}{m_g} \right) \right) M (t, t')
\end{eqnarray}


For the second term in \eqref{seffpaso2} we use
\begin{equation}
\frac{\partial^5}{\partial t^5} \delta (t - t') = -
\frac{\partial^5}{{\partial t'}^5} \delta (t - t')
\end{equation}
and after integrations by parts, discarding terms containing products of deltas with derivatives of Heaviside Theta functions, which are ill defined, and considering that $\xi_q$ and all derivatives of it vanish in the boundaries, we get
\begin{eqnarray}
\int^T_0 d t \int^T_0 d t' \xi^i_{cl} (t) \xi^j_{cl} (t) \theta
(t' - t) \frac{\partial^5}{{\partial t'}^5} \delta (t - t') \tilde{T}^{(0)}_{i
j l m} \xi^l_{cl} (t') \xi^m_q (t') = \theta (0) \int^T_0 d t
\frac{\partial^5 (\xi^i_{cl} (t) \xi^j_{cl} (t))}{\partial t^5}
\tilde{T}^{(0)}_{i j l m} \xi^l_{cl} (t) \xi^m_q (t)
\end{eqnarray}

We proceed similarly for the third term in \eqref{seffpaso2} and for the fourth term we integrate by parts, and we will consider that the terms in the boundary cancel, which is equivalent to assume an adiabatic switching. Then, after removing the terms with deltas and its derivatives, which are treated in the same way as for the other cases, we have
\begin{eqnarray}
& & \int^T_0 d t \int^T_0 d t' \xi^i_{cl} (t) \xi^j_{cl} (t) \theta
(t' - t) \frac{\partial^4 sgn (t' - t) \ln (m_g | t' - t |)}{\partial
t^4} \Xi_{i j l m} \xi^l_{cl} (t') \xi^m_q (t') \nonumber\\
& & = \int^T_0 d t \int^t_0 d t' \frac{\partial^4 \xi^i_{cl} (t')
\xi^j_{cl} (t')}{{\partial t'}^4} \ln (m_g | t - t' |) \Xi_{i j l m}
\xi^l_{cl} (t) \xi^m_q (t)
\end{eqnarray} 

With all that, after integrating by parts in the first and last terms we have that the effective action now is
\begin{eqnarray}
S_{eff} [\xi] &=& - 2 m_0 \int^T_0 d t \frac{\partial^2
\xi_{cl}^i}{\partial t^2} \xi_q^i \nonumber\\
& & - 4 G m_0^2 \theta (0) \int^T_0 d t \frac{\partial^5 (\xi^i_{cl} (t)
\xi^j_{cl} (t))}{\partial t^5} \tilde{T}^{(0)}_{i j l m}
\xi^l_{cl} (t) \xi^m_q (t) \nonumber\\
& & + 4 G m_0^2 m_g^2 \theta (0) \int^T_0 d t \frac{\partial^3 (\xi^i_{cl}
(t) \xi^j_{cl} (t))}{\partial t^3} \tilde{S} _{i j l m}
\xi^l_{cl} (t) \xi^m_q (t) \nonumber\\
& & - 4 G m_0^2 m_g^2 \int^T_0 d t \int^t_0 d t' \frac{\partial^4
\xi^i_{cl} (t') \xi^j_{cl} (t')}{{\partial t'}^4} \ln (m_g | t -
t' |) \Xi_{i j l m} \xi^l_{cl} (t) \xi^m_q (t)
\nonumber\\
& & + i \frac{8 G m_0^2}{\pi} \int^T_0 d t \int^T_0 d t'  \frac{\partial^2
(\xi^i_{cl} (t) \xi^j_q (t))}{\partial t^2} K_{i j l m} (t, t')
\frac{\partial^2 (\xi^l_{cl} (t') \xi^m_q (t'))}{{\partial t'}^2}
\end{eqnarray}

Now we use as prescription $\theta(0)=1/2$ and that is the equation presented in the main text.

\section{Integrals for the noises}
\label{sec:noiseintegrals}

In this appendix we evaluate the integrals $I_0$, $I_1$ and $I_2$ for the vacuum, thermal and squeeze state cases.

\subsection{Vacuum}
For the vacuum case, $A = 1$, $B = 0$, so
\begin{equation}
M (t, t') = \frac{1}{2} \cos (\omega (t - t'))
\end{equation}
Then, we first compute $I_0$:
\begin{equation}
I_0 = \frac{1}{2} \int^{\infty}_0 d \omega \omega \cos (\omega (t - t')) -
\frac{1}{2} \int^{m_g}_0 d \omega \omega \cos (\omega (t - t'))
\end{equation}

The second integral is finite and can be computed: 
\begin{equation}
\int^{m_g}_0 d \omega \omega \cos (\omega (t - t'))=\frac{- 1 + \cos (m_g (t - t')) + m_g (t - t') \sin (m_g (t - t'))}{(t -
t')^2}
\end{equation}

To compute the first integral we need to introduce a regulator $\epsilon>0$, which means that we are dealing with this integral in the distributional sense:
\begin{equation}
\int_0^{\infty} d \omega \omega \cos (\omega (t - t')) \rightarrow \int_0^{\infty} d \omega \omega
e^{- \varepsilon \omega} \cos (\omega (t - t'))
\end{equation}
and after doing all the manipulations we will take the limit $\epsilon\to 0$. Therefore, taking the real part
we have:
\begin{equation}
\int_0^{\infty} d \omega \omega
e^{- \varepsilon \omega} \cos (\omega (t - t')) = \Re \int_0^{\infty} d \omega \omega
e^{\omega (i (t - t') - \varepsilon)} = - \Re\left( \frac{1}{((t - t') + i
\varepsilon)^2} \right)
\end{equation}
and after taking the limit $\epsilon\to0$ and expanding the result of the finite integral up to order $m_g^2$ we get
\begin{equation}
I_0 = - \frac{1}{2} PF \left( \frac{1}{(t - t')^2} \right) - \frac{m_g^2}{4}
\end{equation}
where $PF$ is the Hadamard finite part. 

For $I_1$ the integral can be computed and we obtain
\begin{equation}
I_1 = - \frac{1}{2} Ci (m_g (t - t'))
\end{equation}
Where $Ci$ is the Cosine Integral function. Since $I_1$ is multiplied by $m_g^2$ in $K_{i j l m}$, we expand this result just to order zero in $m_g$ and we get
\begin{equation}
I_1 = - \frac{1}{2} (\gamma + \ln (m_g | t - t' |))
\end{equation}

For $I_2$, we have
\begin{equation}
I_2=\frac{\gamma}{2} \int^{\infty}_{m_g} d \omega \frac{1}{\omega} \cos (\omega
(t - t')) + \frac{1}{2} \int^{\infty}_{m_g} d \omega \frac{1}{\omega} \ln
\left( \frac{\omega}{m_g} \right) \cos (\omega (t - t'))
\end{equation}
The first integral is basically what we got from $I_1$ multiplied by the Euler-Mascheroni constant, and for the second one, doing the change of variables $u = \omega (t - t')$ and $x = m_g (t - t')$ we have
\begin{equation}
\int^{\infty}_{m_g} d \omega \frac{1}{\omega} \ln \left( \frac{\omega}{m_g}
\right) \cos (\omega (t - t')) \rightarrow \int^{\infty}_x d u \frac{1}{u} \ln (u) \cos (u) - \ln (x) \int^{\infty}_x d
u \frac{1}{u} \cos (u)
\end{equation}
Both integrals can be solved in the computational sofwware Mathematica, and since $I_2$ is multiplied by a $m_g^2$ in $K_{i j l m}$, expanding up to order zero in $m_g$ we obtain that
\begin{equation}
\int^{\infty}_{m_g} d \omega \frac{1}{\omega} \ln \left( \frac{\omega}{m_g}
\right) \cos (\omega (t - t')) = \frac{1}{2} \gamma^2 - \frac{1}{24} \pi^2 + \frac{1}{2} (\ln (m_g | t -
t' |))^2 + \gamma \ln (m_g | t - t' |)
\end{equation}
Therefore,
\begin{equation}
I_2 = \frac{1}{4} (\ln (m_g | t - t' |))^2 - \frac{1}{4} \gamma^2 -
\frac{1}{48} \pi^2
\end{equation}
Since the noise enters in the equation of motion only through the second derivative of $\mathcal{N}_{i j}$, constant terms in $I_0$, $I_1$ and $I_2$ do not contribute to that noise correlation, so we will discard them and we will have:
\begin{eqnarray}
I_0 &=& - \frac{1}{2} PF \left( \frac{1}{(t - t')^2} \right)\\
I_1 &=& - \frac{1}{2} \ln (m_g | t - t' |) \\
I_2 &=& \frac{1}{4} (\ln (m_g | t - t' |))^2
\end{eqnarray}


\subsection{Thermal}
In the thermal case, we have that $A = \coth (\beta \omega)$, $B = 2 \csch (\beta \omega)$, so
\begin{eqnarray}
M (t, t') &=& \frac{1}{2 \tanh \left( \frac{\beta \omega}{2} \right)} \cos
(\omega (t - t')) = \frac{1}{2} \left( \frac{2}{e^{\beta \omega} - 1} + 1
\right) \cos (\omega (t - t')) \label{Mtherm}
\end{eqnarray}
With this,
\begin{eqnarray}
\label{initialI0th}
I_0 &=& \frac{1}{2} \int^{\infty}_0 d \omega \omega \cos (\omega (t - t')) -
\frac{1}{2} \int^{m_g}_0 d \omega \omega \cos (\omega (t - t')) \nonumber\\
& & + \int^{\infty}_0 d \omega \frac{\omega \cos (\omega (t - t'))}{e^{\beta \omega}
- 1} - \int^{m_g}_0 d \omega \frac{\omega \cos (\omega (t - t'))}{e^{\beta
\omega} - 1}
\end{eqnarray}

We observe that the first line in \eqref{initialI0th} will give the same integrals computed for the vacuum case, so we focus here only in the second line. The last term in the second line is finite, and that full integral computed in the software Mathematica is composed by Hurwitz Zeta functions and Hypergeometric functions, but expanding up to order $m_g^2$ is simply:
\begin{equation}
\int^{m_g}_0 d \omega \frac{\omega \cos (\omega (t - t'))}{e^{\beta \omega} -
1} \approx \frac{m_g}{\beta} - \frac{m_g^2}{4}
\end{equation}

To solve the first integral in the second line of \eqref{initialI0th} we start from the following integral:
\begin{equation}
\int_0^{\infty} d \omega \frac{\sin (\omega (t - t'))}{e^{\beta \omega} - 1} = \int_0^{\infty}
d \omega \sin (\omega (t - t')) \underset{n = 1}{\overset{\infty}{\sum}} e^{- \beta n \omega}
\end{equation}
after interchanging integral and sum, the integration gives
\begin{equation}
\int_0^{\infty} d \omega \frac{\sin (\omega (t - t'))}{e^{\beta \omega} - 1} = - \frac{1}{2 (t - t')} + \frac{\pi}{2\beta} \coth (\frac{\pi}{\beta}(t - t'))
\end{equation}
With this integral, we can derive respect to $t-t'$ to get
\begin{equation}
\int_0^{\infty} d \omega \frac{\omega \cos (\omega (t - t'))}{e^{\beta\omega} - 1} = \frac{1}{2(t - t')^2} - \frac{\pi^2}{2\beta^2(\sinh (\frac{\pi}{\beta}(t - t')))^2}
\end{equation}
We notice that the first term is cancelled by the vacuum term coming from the first term in first line of \eqref{initialI0th}. Thus, putting all the pieces together
\begin{equation}
I_0 = - \frac{\pi^2}{2 \beta^2 \left( \sinh \left( \frac{\pi}{\beta} (t - t')
\right) \right)^2} - \frac{m_g}{\beta}
\end{equation}

For $I_1$ we have
\begin{equation}
I_1 = \int^{\infty}_{m_g} d \omega \frac{1}{\omega} \frac{1}{e^{\beta \omega}
- 1} \cos (\omega (t - t')) + \frac{1}{2} \int^{\infty}_{m_g} d \omega
\frac{1}{\omega} \cos (\omega (t - t'))
\end{equation}
This can be rewritten as
\begin{equation}
I_1 = \frac{1}{2} \int^{\infty}_{m_g} d \omega \frac{1}{\omega} \coth \left(
\frac{1}{2} \beta \omega \right) \cos (\omega (t - t'))
\end{equation}
Using the identity:
\begin{equation}
\label{idecoth}
\coth (x) = \frac{1}{x} + \underset{n = 1}{\overset{\infty}{\sum}} \frac{2
x}{x^2 + \pi^2 n^2}
\end{equation}
it can be arranged as
\begin{equation}
I_1 = \frac{1}{\beta} \int^{\infty}_{m_g} d \omega \frac{1}{\omega^2} \cos
(\omega (t - t')) + \frac{2}{\beta} \underset{n = 1}{\overset{\infty}{\sum}}
\int^{\infty}_0 d \omega \frac{\cos (\omega (t - t'))}{\omega^2 + \frac{4
\pi^2 n^2}{\beta^2}} - \frac{2}{\beta} \underset{n =
1}{\overset{\infty}{\sum}} \int^{m_g}_0 d \omega \frac{\cos (\omega (t -
t'))}{\omega^2 + \frac{4 \pi^2 n^2}{\beta^2}}
\end{equation}
All those integrals can be done in Mathematica, giving terms that contain Sine Integrals and Exponential Integral functions. In any case, expanding up to order zero in $m_g$, as $I_1$ is multiplied by $m_g^2$, we get that
\begin{equation}
I_1 = - \frac{\pi}{2 \beta} | t - t' | + \frac{1}{\beta m_g} + \underset{n =
1}{\overset{\infty}{\sum}} \frac{1}{2 n} e^{- \frac{2 \pi n}{\beta} | t - t'
|}
\end{equation}
The sum corresponds to a logarithm and after arranging we obtain
\begin{equation}
I_1 = \frac{1}{\beta m_g} - \frac{1}{2} \ln \left( 2 \sinh \left(
\frac{\pi}{\beta} | t - t' | \right) \right)
\end{equation}

To compute $I_2$ we have that
\begin{equation}
I_2 = \gamma I_1 + \int^{\infty}_{m_g} d \omega \frac{1}{\omega} \ln \left(
\frac{\omega}{m_g} \right) M (t, t')
\end{equation}
and it can be arranged as
\begin{eqnarray}
I_2 = \frac{\gamma}{\beta m_g} - \frac{\gamma}{2} \ln \left( 2 \sinh \left(
\frac{\pi}{\beta} | t - t' | \right) \right) + \frac{1}{2} \int^{\infty}_{m_g}
d \omega \frac{1}{\omega} \ln \left( \frac{\omega}{m_g} \right) \coth \left(
\frac{1}{2} \beta \omega \right) \cos (\omega (t - t'))
\end{eqnarray}
Using the identity \eqref{idecoth} we can arrange ths expression as
\begin{eqnarray}
I_2 &=& \frac{\gamma}{\beta m_g} - \frac{\gamma}{2} \ln \left( 2 \sinh \left(
\frac{\pi}{\beta} | t - t' | \right) \right) + \frac{1}{\beta}
\int^{\infty}_{m_g} d \omega \frac{1}{\omega^2} \ln \left( \frac{\omega}{m_g}
\right) \cos (\omega (t - t')) \nonumber\\
& & + \frac{2}{\beta} \int^{\infty}_0 d \omega
\underset{n = 1}{\overset{\infty}{\sum}} \frac{1}{\omega^2 + \frac{4 \pi^2
n^2}{\beta^2}} \ln \left( \frac{\omega}{m_g} \right) \cos (\omega (t - t')) \nonumber\\
& & - \frac{2}{\beta} \int^{m_g}_0 d \omega \underset{n = 1}{\overset{\infty}{\sum}}
\frac{1}{\omega^2 + \frac{4 \pi^2 n^2}{\beta^2}} \ln \left( \frac{\omega}{m_g}
\right) \cos (\omega (t - t'))
\end{eqnarray}

The last term will give only an object that expanded in series starts at order $m_g$. Since $I_2$ is multiplied by $m_g^2$ we can ignore that term. The other integrals can be done in Mathematica and expanded in series up to order zero in $m_g$ and with that we get
\begin{eqnarray}
I_2 &=& \frac{\gamma + 1}{\beta m_g} - \frac{\gamma}{2} \ln \left( 2 \sinh
\left( \frac{\pi}{\beta} | t - t' | \right) \right) - \frac{\pi}{2 \beta} | t
- t' | (1 - \gamma - \ln (m_g | t - t' |)) \nonumber\\
& & + \underset{n =
1}{\overset{\infty}{\sum}} \frac{1}{2 n} \ln \left( \frac{2 \pi n}{\beta m_g}
\right) e^{- \frac{2 \pi n}{\beta} | t - t' |} \nonumber\\
& & + \underset{n =
1}{\overset{\infty}{\sum}} \frac{1}{4 n} \left( e^{\frac{2 \pi n}{\beta} | t -
t' |} Ei \left( - \frac{2 \pi n}{\beta} | t - t' | \right) - e^{-
\frac{2 \pi n}{\beta} | t - t' |} Ei \left( \frac{2 \pi n}{\beta} | t -
t' | \right) \right)
\end{eqnarray}
where $Ei$ is the Exponential Integral function, and those series cannot be expressed in a closed form.

As in the vacuum case, we remove the constant terms and we have:
\begin{eqnarray}
I_0 &=& - \frac{\pi^2}{2 \beta^2 \left( \sinh \left( \frac{\pi}{\beta} (t - t')
\right) \right)^2} \\
I_1 &=& - \frac{1}{2} \ln \left( 2 \sinh \left( \frac{\pi}{\beta} | t - t' |
\right) \right) \\
I_2 &=& - \frac{\gamma}{2} \ln \left( 2 \sinh \left( \frac{\pi}{\beta} | t - t'
| \right) \right) - \frac{\pi}{2 \beta} | t - t' | (1 - \gamma - \ln (m_g | t - t' |)) \nonumber\\
& & + \underset{n = 1}{\overset{\infty}{\sum}} \frac{1}{2 n} \ln \left(
\frac{2 \pi n}{\beta m_g} \right) e^{- \frac{2 \pi n}{\beta} | t - t' |} \nonumber\\
& & + \underset{n = 1}{\overset{\infty}{\sum}} \frac{1}{4 n} \left( e^{\frac{2 \pi
n}{\beta} | t - t' |} Ei \left( - \frac{2 \pi n}{\beta} | t - t' |
\right) - e^{- \frac{2 \pi n}{\beta} | t - t' |} Ei \left( \frac{2 \pi
n}{\beta} | t - t' | \right) \right)
\end{eqnarray}

\subsection{Squeeze}
For the squeeze state, $A = \frac{\cosh r + e^{i \phi} \sinh r}{\cosh r - e^{i \phi} \sinh r}$, $B =
0$, so we have that
\begin{eqnarray}
\Re (A) &=& \frac{1}{\cosh (2 r) - \cos \phi \sinh (2 r)} \\ \Im (A) &=& \frac{\sin \phi \sinh (2 r)}{\cosh (2 r) - \cos \phi \sinh (2 r)} \\
| A |^2 &=& \frac{\cosh (2 r) + \cos \phi \sinh (2 r)}{\cosh (2 r) - \cos \phi \sinh
(2 r)}
\end{eqnarray}
and hence,
\begin{eqnarray}
M (t, t') &=& \frac{1}{2} \cosh (2 r) \cos (\omega (t - t')) - \frac{1}{2}
\cos \phi \sinh (2 r) \cos (\omega (t + t')) - \frac{1}{2} \sin \phi \sinh (2
r) \sin (\omega (t + t'))
\end{eqnarray}


We can see that the first term can be treated as in the vacuum case multiplied by $\cosh{(2r)}$, while the second term produces the same integrals with $t-t'$ replaced by $t+t'$ and multiplied by $-\cos{\phi}\sinh{(2r)}$. So, we have to compute only the integrals with $\sinh{(\omega(t+t'))}$. The first integral is
\begin{equation}
\int^{\infty}_{m_g} d \omega \omega \sin (\omega (t + t')) = \int^{\infty}_0
d \omega \omega \sin (\omega (t + t')) - \int^{m_g}_0 d \omega \omega \sin
(\omega (t + t'))
\end{equation}
The second term is just a finite part that starts at order $m_g^3$, and the first one must be understood in the distributional sense giving
\begin{equation}
\int^{\infty}_{m_g} d \omega \omega \sin (\omega (t + t')) \approx - \pi
\delta' (t + t')
\end{equation}
Proceeding similarly for the other integrals with sine, where the split term that goes from $0$ to $m_g$ is finite and since those integrals are multiplied by a factor of $m_g^2$ we ignore the terms of higher order and we will have
\begin{eqnarray}
\int^{\infty}_{m_g} d \omega \frac{1}{\omega} \sin (\omega (t + t')) &\approx& \frac{\pi}{2} \\
\int^{\infty}_{m_g} d \omega \frac{1}{\omega} \left( \gamma + \ln \left(
\frac{\omega}{m_g} \right) \right) \sin (\omega (t + t')) &\approx& -
\frac{\pi}{2} \ln (m_g | t + t' |)
\end{eqnarray}

With all that, the terms $I_0$, $I_1$ and $I_2$ are

\begin{eqnarray}
I_0 &=& - \frac{1}{2} \cosh (2 r) PF \left( \frac{1}{(t - t')^2} \right)
+ \frac{1}{2} \cos \phi \sinh (2 r) PF \left( \frac{1}{(t + t')^2}
\right) - \frac{m_g^2}{4} (\cosh (2 r) - \cos \phi \sinh (2 r)) \nonumber\\
& & + \frac{\pi}{2} \sin \phi \sinh (2 r) \delta' (t + t') \\
I_1 &=& - \frac{1}{2} \cosh (2 r) (\gamma + \ln (m_g | t - t' |)) + \frac{1}{2}
\cos \phi \sinh (2 r) (\gamma + \ln (m_g | t + t' |)) - \frac{\pi}{4} \sin
\phi \sinh (2 r) \\
I_2 &=& \frac{1}{4} \cosh (2 r) (\ln (m_g | t - t' |))^2 - \frac{1}{4} \cos
\phi \sinh (2 r) (\ln (m_g | t + t' |))^2 - \frac{1}{4} \left( \gamma^2 +
\frac{\pi^2}{12} \right) (\cosh (2 r) - \cos \phi \sinh (2 r)) \nonumber\\
& & + \frac{\pi}{4}
\sin \phi \sinh (2 r) \ln (m_g | t + t' |)
\end{eqnarray}

Removing the constant term as in the previous cases:
\begin{eqnarray}
I_0 &=& - \frac{1}{2} \cosh (2 r) PF \left( \frac{1}{(t - t')^2} \right)
+ \frac{1}{2} \cos \phi \sinh (2 r) PF \left( \frac{1}{(t + t')^2}
\right) + \frac{\pi}{2} \sin \phi \sinh (2 r) \delta' (t + t') \\
I_1 &=& - \frac{1}{2} \cosh (2 r) \ln (m_g | t - t' |) + \frac{1}{2} \cos \phi
\sinh (2 r) \ln (m_g | t + t' |) \\
I_2 &=& \frac{1}{4} \cosh (2 r) (\ln (m_g | t - t' |))^2 - \frac{1}{4} \cos
\phi \sinh (2 r) (\ln (m_g | t + t' |))^2 + \frac{\pi}{4} \sin \phi \sinh (2
r) \ln (m_g | t + t' |)
\end{eqnarray}

\bibliography{RefsGWVSR}

\begin{thebibliography}{37}%
\makeatletter
\providecommand \@ifxundefined [1]{%
 \@ifx{#1\undefined}
}%
\providecommand \@ifnum [1]{%
 \ifnum #1\expandafter \@firstoftwo
 \else \expandafter \@secondoftwo
 \fi
}%
\providecommand \@ifx [1]{%
 \ifx #1\expandafter \@firstoftwo
 \else \expandafter \@secondoftwo
 \fi
}%
\providecommand \natexlab [1]{#1}%
\providecommand \enquote  [1]{``#1''}%
\providecommand \bibnamefont  [1]{#1}%
\providecommand \bibfnamefont [1]{#1}%
\providecommand \citenamefont [1]{#1}%
\providecommand \href@noop [0]{\@secondoftwo}%
\providecommand \href [0]{\begingroup \@sanitize@url \@href}%
\providecommand \@href[1]{\@@startlink{#1}\@@href}%
\providecommand \@@href[1]{\endgroup#1\@@endlink}%
\providecommand \@sanitize@url [0]{\catcode `\\12\catcode `\$12\catcode
  `\&12\catcode `\#12\catcode `\^12\catcode `\_12\catcode `\%12\relax}%
\providecommand \@@startlink[1]{}%
\providecommand \@@endlink[0]{}%
\providecommand \url  [0]{\begingroup\@sanitize@url \@url }%
\providecommand \@url [1]{\endgroup\@href {#1}{\urlprefix }}%
\providecommand \urlprefix  [0]{URL }%
\providecommand \Eprint [0]{\href }%
\providecommand \doibase [0]{https://doi.org/}%
\providecommand \selectlanguage [0]{\@gobble}%
\providecommand \bibinfo  [0]{\@secondoftwo}%
\providecommand \bibfield  [0]{\@secondoftwo}%
\providecommand \translation [1]{[#1]}%
\providecommand \BibitemOpen [0]{}%
\providecommand \bibitemStop [0]{}%
\providecommand \bibitemNoStop [0]{.\EOS\space}%
\providecommand \EOS [0]{\spacefactor3000\relax}%
\providecommand \BibitemShut  [1]{\csname bibitem#1\endcsname}%
\let\auto@bib@innerbib\@empty
\bibitem [{\citenamefont {Abbott}\ \emph {et~al.}(2016)\citenamefont {Abbott}
  \emph {et~al.}}]{LIGOScientific:2016aoc}%
  \BibitemOpen
  \bibfield  {author} {\bibinfo {author} {\bibfnamefont {B.~P.}\ \bibnamefont
  {Abbott}} \emph {et~al.} (\bibinfo {collaboration} {LIGO Scientific,
  Virgo}),\ }\href {https://doi.org/10.1103/PhysRevLett.116.061102} {\bibfield
  {journal} {\bibinfo  {journal} {Phys. Rev. Lett.}\ }\textbf {\bibinfo
  {volume} {116}},\ \bibinfo {pages} {061102} (\bibinfo {year} {2016})},\
  \Eprint {https://arxiv.org/abs/1602.03837} {arXiv:1602.03837 [gr-qc]}
  \BibitemShut {NoStop}%
\bibitem [{\citenamefont {Parikh}\ \emph {et~al.}(2021)\citenamefont {Parikh},
  \citenamefont {Wilczek},\ and\ \citenamefont
  {Zahariade}}]{PhysRevD.104.046021}%
  \BibitemOpen
  \bibfield  {author} {\bibinfo {author} {\bibfnamefont {M.}~\bibnamefont
  {Parikh}}, \bibinfo {author} {\bibfnamefont {F.}~\bibnamefont {Wilczek}},\
  and\ \bibinfo {author} {\bibfnamefont {G.}~\bibnamefont {Zahariade}},\ }\href
  {https://doi.org/10.1103/PhysRevD.104.046021} {\bibfield  {journal} {\bibinfo
   {journal} {Phys. Rev. D}\ }\textbf {\bibinfo {volume} {104}},\ \bibinfo
  {pages} {046021} (\bibinfo {year} {2021})}\BibitemShut {NoStop}%
\bibitem [{\citenamefont {Fierz}\ and\ \citenamefont
  {Pauli}(1939)}]{Fierz:1939ix}%
  \BibitemOpen
  \bibfield  {author} {\bibinfo {author} {\bibfnamefont {M.}~\bibnamefont
  {Fierz}}\ and\ \bibinfo {author} {\bibfnamefont {W.}~\bibnamefont {Pauli}},\
  }\href {https://doi.org/10.1098/rspa.1939.0140} {\bibfield  {journal}
  {\bibinfo  {journal} {Proc. Roy. Soc. Lond. A}\ }\textbf {\bibinfo {volume}
  {173}},\ \bibinfo {pages} {211} (\bibinfo {year} {1939})}\BibitemShut
  {NoStop}%
\bibitem [{\citenamefont {van Dam}\ and\ \citenamefont
  {Veltman}(1970)}]{vanDam:1970vg}%
  \BibitemOpen
  \bibfield  {author} {\bibinfo {author} {\bibfnamefont {H.}~\bibnamefont {van
  Dam}}\ and\ \bibinfo {author} {\bibfnamefont {M.~J.~G.}\ \bibnamefont
  {Veltman}},\ }\href {https://doi.org/10.1016/0550-3213(70)90416-5} {\bibfield
   {journal} {\bibinfo  {journal} {Nucl. Phys. B}\ }\textbf {\bibinfo {volume}
  {22}},\ \bibinfo {pages} {397} (\bibinfo {year} {1970})}\BibitemShut
  {NoStop}%
\bibitem [{\citenamefont {Zakharov}(1970)}]{Zakharov:1970cc}%
  \BibitemOpen
  \bibfield  {author} {\bibinfo {author} {\bibfnamefont {V.~I.}\ \bibnamefont
  {Zakharov}},\ }\href@noop {} {\bibfield  {journal} {\bibinfo  {journal} {JETP
  Lett.}\ }\textbf {\bibinfo {volume} {12}},\ \bibinfo {pages} {312} (\bibinfo
  {year} {1970})}\BibitemShut {NoStop}%
\bibitem [{\citenamefont {Boulware}\ and\ \citenamefont
  {Deser}(1972)}]{Boulware:1972yco}%
  \BibitemOpen
  \bibfield  {author} {\bibinfo {author} {\bibfnamefont {D.~G.}\ \bibnamefont
  {Boulware}}\ and\ \bibinfo {author} {\bibfnamefont {S.}~\bibnamefont
  {Deser}},\ }\href {https://doi.org/10.1103/PhysRevD.6.3368} {\bibfield
  {journal} {\bibinfo  {journal} {Phys. Rev. D}\ }\textbf {\bibinfo {volume}
  {6}},\ \bibinfo {pages} {3368} (\bibinfo {year} {1972})}\BibitemShut
  {NoStop}%
\bibitem [{\citenamefont {Vainshtein}(1972)}]{Vainshtein:1972sx}%
  \BibitemOpen
  \bibfield  {author} {\bibinfo {author} {\bibfnamefont {A.~I.}\ \bibnamefont
  {Vainshtein}},\ }\href {https://doi.org/10.1016/0370-2693(72)90147-5}
  {\bibfield  {journal} {\bibinfo  {journal} {Phys. Lett. B}\ }\textbf
  {\bibinfo {volume} {39}},\ \bibinfo {pages} {393} (\bibinfo {year}
  {1972})}\BibitemShut {NoStop}%
\bibitem [{\citenamefont {Dvali}\ \emph
  {et~al.}(2000{\natexlab{a}})\citenamefont {Dvali}, \citenamefont
  {Gabadadze},\ and\ \citenamefont {Porrati}}]{Dvali:2000hr}%
  \BibitemOpen
  \bibfield  {author} {\bibinfo {author} {\bibfnamefont {G.~R.}\ \bibnamefont
  {Dvali}}, \bibinfo {author} {\bibfnamefont {G.}~\bibnamefont {Gabadadze}},\
  and\ \bibinfo {author} {\bibfnamefont {M.}~\bibnamefont {Porrati}},\ }\href
  {https://doi.org/10.1016/S0370-2693(00)00669-9} {\bibfield  {journal}
  {\bibinfo  {journal} {Phys. Lett. B}\ }\textbf {\bibinfo {volume} {485}},\
  \bibinfo {pages} {208} (\bibinfo {year} {2000}{\natexlab{a}})},\ \Eprint
  {https://arxiv.org/abs/hep-th/0005016} {arXiv:hep-th/0005016} \BibitemShut
  {NoStop}%
\bibitem [{\citenamefont {Dvali}\ \emph
  {et~al.}(2000{\natexlab{b}})\citenamefont {Dvali}, \citenamefont
  {Gabadadze},\ and\ \citenamefont {Porrati}}]{Dvali:2000rv}%
  \BibitemOpen
  \bibfield  {author} {\bibinfo {author} {\bibfnamefont {G.~R.}\ \bibnamefont
  {Dvali}}, \bibinfo {author} {\bibfnamefont {G.}~\bibnamefont {Gabadadze}},\
  and\ \bibinfo {author} {\bibfnamefont {M.}~\bibnamefont {Porrati}},\ }\href
  {https://doi.org/10.1016/S0370-2693(00)00631-6} {\bibfield  {journal}
  {\bibinfo  {journal} {Phys. Lett. B}\ }\textbf {\bibinfo {volume} {484}},\
  \bibinfo {pages} {112} (\bibinfo {year} {2000}{\natexlab{b}})},\ \Eprint
  {https://arxiv.org/abs/hep-th/0002190} {arXiv:hep-th/0002190} \BibitemShut
  {NoStop}%
\bibitem [{\citenamefont {de~Rham}\ \emph {et~al.}(2011)\citenamefont
  {de~Rham}, \citenamefont {Gabadadze},\ and\ \citenamefont
  {Tolley}}]{deRham:2010kj}%
  \BibitemOpen
  \bibfield  {author} {\bibinfo {author} {\bibfnamefont {C.}~\bibnamefont
  {de~Rham}}, \bibinfo {author} {\bibfnamefont {G.}~\bibnamefont {Gabadadze}},\
  and\ \bibinfo {author} {\bibfnamefont {A.~J.}\ \bibnamefont {Tolley}},\
  }\href {https://doi.org/10.1103/PhysRevLett.106.231101} {\bibfield  {journal}
  {\bibinfo  {journal} {Phys. Rev. Lett.}\ }\textbf {\bibinfo {volume} {106}},\
  \bibinfo {pages} {231101} (\bibinfo {year} {2011})},\ \Eprint
  {https://arxiv.org/abs/1011.1232} {arXiv:1011.1232 [hep-th]} \BibitemShut
  {NoStop}%
\bibitem [{\citenamefont {Cohen}\ and\ \citenamefont
  {Glashow}(2006{\natexlab{a}})}]{vsr1}%
  \BibitemOpen
  \bibfield  {author} {\bibinfo {author} {\bibfnamefont {A.~G.}\ \bibnamefont
  {Cohen}}\ and\ \bibinfo {author} {\bibfnamefont {S.~L.}\ \bibnamefont
  {Glashow}},\ }\href {https://doi.org/10.1103/PhysRevLett.97.021601}
  {\bibfield  {journal} {\bibinfo  {journal} {Phys. Rev. Lett.}\ }\textbf
  {\bibinfo {volume} {97}},\ \bibinfo {pages} {021601} (\bibinfo {year}
  {2006}{\natexlab{a}})},\ \Eprint {https://arxiv.org/abs/hep-ph/0601236}
  {arXiv:hep-ph/0601236} \BibitemShut {NoStop}%
\bibitem [{\citenamefont {Cohen}\ and\ \citenamefont
  {Glashow}(2006{\natexlab{b}})}]{vsr2}%
  \BibitemOpen
  \bibfield  {author} {\bibinfo {author} {\bibfnamefont {A.~G.}\ \bibnamefont
  {Cohen}}\ and\ \bibinfo {author} {\bibfnamefont {S.~L.}\ \bibnamefont
  {Glashow}},\ }\href@noop {} {\  (\bibinfo {year} {2006}{\natexlab{b}})},\
  \Eprint {https://arxiv.org/abs/hep-ph/0605036} {arXiv:hep-ph/0605036}
  \BibitemShut {NoStop}%
\bibitem [{\citenamefont {Alfaro}\ and\ \citenamefont
  {Rivelles}(2013)}]{Alfaro:2013uva}%
  \BibitemOpen
  \bibfield  {author} {\bibinfo {author} {\bibfnamefont {J.}~\bibnamefont
  {Alfaro}}\ and\ \bibinfo {author} {\bibfnamefont {V.~O.}\ \bibnamefont
  {Rivelles}},\ }\href {https://doi.org/10.1103/PhysRevD.88.085023} {\bibfield
  {journal} {\bibinfo  {journal} {Phys. Rev. D}\ }\textbf {\bibinfo {volume}
  {88}},\ \bibinfo {pages} {085023} (\bibinfo {year} {2013})},\ \Eprint
  {https://arxiv.org/abs/1305.1577} {arXiv:1305.1577 [hep-th]} \BibitemShut
  {NoStop}%
\bibitem [{\citenamefont {Alfaro}\ \emph {et~al.}(2015)\citenamefont {Alfaro},
  \citenamefont {Gonz\'alez},\ and\ \citenamefont
  {\'Avila}}]{PhysRevD.91.105007}%
  \BibitemOpen
  \bibfield  {author} {\bibinfo {author} {\bibfnamefont {J.}~\bibnamefont
  {Alfaro}}, \bibinfo {author} {\bibfnamefont {P.}~\bibnamefont {Gonz\'alez}},\
  and\ \bibinfo {author} {\bibfnamefont {R.}~\bibnamefont {\'Avila}},\ }\href
  {https://doi.org/10.1103/PhysRevD.91.105007} {\bibfield  {journal} {\bibinfo
  {journal} {Phys. Rev. D}\ }\textbf {\bibinfo {volume} {91}},\ \bibinfo
  {pages} {105007} (\bibinfo {year} {2015})}\BibitemShut {NoStop}%
\bibitem [{\citenamefont {Cheon}\ \emph {et~al.}(2009)\citenamefont {Cheon},
  \citenamefont {Lee},\ and\ \citenamefont {Lee}}]{Cheon:2009zx}%
  \BibitemOpen
  \bibfield  {author} {\bibinfo {author} {\bibfnamefont {S.}~\bibnamefont
  {Cheon}}, \bibinfo {author} {\bibfnamefont {C.}~\bibnamefont {Lee}},\ and\
  \bibinfo {author} {\bibfnamefont {S.~J.}\ \bibnamefont {Lee}},\ }\href
  {https://doi.org/10.1016/j.physletb.2009.07.007} {\bibfield  {journal}
  {\bibinfo  {journal} {Phys. Lett. B}\ }\textbf {\bibinfo {volume} {679}},\
  \bibinfo {pages} {73} (\bibinfo {year} {2009})},\ \Eprint
  {https://arxiv.org/abs/0904.2065} {arXiv:0904.2065 [hep-th]} \BibitemShut
  {NoStop}%
\bibitem [{\citenamefont {Alfaro}\ and\ \citenamefont
  {Soto}(2019)}]{PhysRevD.100.055029}%
  \BibitemOpen
  \bibfield  {author} {\bibinfo {author} {\bibfnamefont {J.}~\bibnamefont
  {Alfaro}}\ and\ \bibinfo {author} {\bibfnamefont {A.}~\bibnamefont {Soto}},\
  }\href {https://doi.org/10.1103/PhysRevD.100.055029} {\bibfield  {journal}
  {\bibinfo  {journal} {Phys. Rev. D}\ }\textbf {\bibinfo {volume} {100}},\
  \bibinfo {pages} {055029} (\bibinfo {year} {2019})}\BibitemShut {NoStop}%
\bibitem [{\citenamefont {Maluf}\ and\ \citenamefont
  {Olmo}(2023)}]{Maluf:2023gbf}%
  \BibitemOpen
  \bibfield  {author} {\bibinfo {author} {\bibfnamefont {R.~V.}\ \bibnamefont
  {Maluf}}\ and\ \bibinfo {author} {\bibfnamefont {G.~J.}\ \bibnamefont
  {Olmo}},\ }\href {https://doi.org/10.1103/PhysRevD.108.095022} {\bibfield
  {journal} {\bibinfo  {journal} {Phys. Rev. D}\ }\textbf {\bibinfo {volume}
  {108}},\ \bibinfo {pages} {095022} (\bibinfo {year} {2023})},\ \Eprint
  {https://arxiv.org/abs/2306.14309} {arXiv:2306.14309 [hep-th]} \BibitemShut
  {NoStop}%
\bibitem [{\citenamefont {Belchior}\ and\ \citenamefont
  {Maluf}(2024)}]{Belchior:2024eyz}%
  \BibitemOpen
  \bibfield  {author} {\bibinfo {author} {\bibfnamefont {F.~M.}\ \bibnamefont
  {Belchior}}\ and\ \bibinfo {author} {\bibfnamefont {R.~V.}\ \bibnamefont
  {Maluf}},\ }\href {https://doi.org/10.1016/j.physletb.2024.138794} {\bibfield
   {journal} {\bibinfo  {journal} {Phys. Lett. B}\ }\textbf {\bibinfo {volume}
  {855}},\ \bibinfo {pages} {138794} (\bibinfo {year} {2024})}\BibitemShut
  {NoStop}%
\bibitem [{\citenamefont {Belchior}\ and\ \citenamefont
  {Maluf}(2026)}]{Belchior:2026feh}%
  \BibitemOpen
  \bibfield  {author} {\bibinfo {author} {\bibfnamefont {F.~M.}\ \bibnamefont
  {Belchior}}\ and\ \bibinfo {author} {\bibfnamefont {R.~V.}\ \bibnamefont
  {Maluf}},\ }\href@noop {} {\  (\bibinfo {year} {2026})},\ \Eprint
  {https://arxiv.org/abs/2601.18621} {arXiv:2601.18621 [hep-th]} \BibitemShut
  {NoStop}%
\bibitem [{\citenamefont {Alfaro}\ and\ \citenamefont
  {Santoni}(2022)}]{Alfaro:2022vpt}%
  \BibitemOpen
  \bibfield  {author} {\bibinfo {author} {\bibfnamefont {J.}~\bibnamefont
  {Alfaro}}\ and\ \bibinfo {author} {\bibfnamefont {A.}~\bibnamefont
  {Santoni}},\ }\href {https://doi.org/10.1016/j.physletb.2022.137080}
  {\bibfield  {journal} {\bibinfo  {journal} {Phys. Lett. B}\ }\textbf
  {\bibinfo {volume} {829}},\ \bibinfo {pages} {137080} (\bibinfo {year}
  {2022})},\ \Eprint {https://arxiv.org/abs/2204.05485} {arXiv:2204.05485
  [gr-qc]} \BibitemShut {NoStop}%
\bibitem [{\citenamefont {Santoni}\ \emph {et~al.}(2023)\citenamefont
  {Santoni}, \citenamefont {Alfaro},\ and\ \citenamefont
  {Soto}}]{Santoni:2023uko}%
  \BibitemOpen
  \bibfield  {author} {\bibinfo {author} {\bibfnamefont {A.}~\bibnamefont
  {Santoni}}, \bibinfo {author} {\bibfnamefont {J.}~\bibnamefont {Alfaro}},\
  and\ \bibinfo {author} {\bibfnamefont {A.}~\bibnamefont {Soto}},\ }\href
  {https://doi.org/10.1103/PhysRevD.108.044072} {\bibfield  {journal} {\bibinfo
   {journal} {Phys. Rev. D}\ }\textbf {\bibinfo {volume} {108}},\ \bibinfo
  {pages} {044072} (\bibinfo {year} {2023})},\ \Eprint
  {https://arxiv.org/abs/2306.02464} {arXiv:2306.02464 [gr-qc]} \BibitemShut
  {NoStop}%
\bibitem [{\citenamefont {Bonilla}\ \emph {et~al.}(2025)\citenamefont
  {Bonilla}, \citenamefont {Santoni}, \citenamefont {Nunes},\ and\
  \citenamefont {Levi~Said}}]{Bonilla:2025mrt}%
  \BibitemOpen
  \bibfield  {author} {\bibinfo {author} {\bibfnamefont {A.}~\bibnamefont
  {Bonilla}}, \bibinfo {author} {\bibfnamefont {A.}~\bibnamefont {Santoni}},
  \bibinfo {author} {\bibfnamefont {R.~C.}\ \bibnamefont {Nunes}},\ and\
  \bibinfo {author} {\bibfnamefont {J.}~\bibnamefont {Levi~Said}},\ }\href
  {https://doi.org/10.1016/j.physletb.2025.139388} {\bibfield  {journal}
  {\bibinfo  {journal} {Phys. Lett. B}\ }\textbf {\bibinfo {volume} {864}},\
  \bibinfo {pages} {139388} (\bibinfo {year} {2025})},\ \Eprint
  {https://arxiv.org/abs/2503.12195} {arXiv:2503.12195 [gr-qc]} \BibitemShut
  {NoStop}%
\bibitem [{\citenamefont {Schwinger}(1961)}]{Schwinger:1960qe}%
  \BibitemOpen
  \bibfield  {author} {\bibinfo {author} {\bibfnamefont {J.~S.}\ \bibnamefont
  {Schwinger}},\ }\href {https://doi.org/10.1063/1.1703727} {\bibfield
  {journal} {\bibinfo  {journal} {J. Math. Phys.}\ }\textbf {\bibinfo {volume}
  {2}},\ \bibinfo {pages} {407} (\bibinfo {year} {1961})}\BibitemShut {NoStop}%
\bibitem [{\citenamefont {Keldysh}(1965)}]{Keldysh:1964ud}%
  \BibitemOpen
  \bibfield  {author} {\bibinfo {author} {\bibfnamefont {L.~V.}\ \bibnamefont
  {Keldysh}},\ }\href {https://doi.org/10.1142/9789811279461_0007} {\bibfield
  {journal} {\bibinfo  {journal} {Sov. Phys. JETP}\ }\textbf {\bibinfo {volume}
  {20}},\ \bibinfo {pages} {1018} (\bibinfo {year} {1965})}\BibitemShut
  {NoStop}%
\bibitem [{\citenamefont {Kamenev}(2011)}]{kamenev_2011}%
  \BibitemOpen
  \bibfield  {author} {\bibinfo {author} {\bibfnamefont {A.}~\bibnamefont
  {Kamenev}},\ }\href {https://doi.org/10.1017/CBO9781139003667} {\emph
  {\bibinfo {title} {Field Theory of Non-Equilibrium Systems}}}\ (\bibinfo
  {publisher} {Cambridge University Press},\ \bibinfo {year}
  {2011})\BibitemShut {NoStop}%
\bibitem [{\citenamefont {Calzetta}\ and\ \citenamefont
  {Hu}(2009)}]{Calzetta:2008iqa}%
  \BibitemOpen
  \bibfield  {author} {\bibinfo {author} {\bibfnamefont {E.~A.}\ \bibnamefont
  {Calzetta}}\ and\ \bibinfo {author} {\bibfnamefont {B.-L.~B.}\ \bibnamefont
  {Hu}},\ }\href {https://doi.org/10.1017/9781009290036} {\emph {\bibinfo
  {title} {{Nonequilibrium Quantum Field Theory}}}}\ (\bibinfo  {publisher}
  {Oxford University Press},\ \bibinfo {year} {2009})\BibitemShut {NoStop}%
\bibitem [{\citenamefont {Rammer}(2007)}]{Rammer:2007zz}%
  \BibitemOpen
  \bibfield  {author} {\bibinfo {author} {\bibfnamefont {J.}~\bibnamefont
  {Rammer}},\ }\href@noop {} {\emph {\bibinfo {title} {{Quantum field theory of
  non-equilibrium states}}}}\ (\bibinfo {year} {2007})\BibitemShut {NoStop}%
\bibitem [{\citenamefont {Feynman}\ and\ \citenamefont
  {Vernon}(1963)}]{FeynmanVernon}%
  \BibitemOpen
  \bibfield  {author} {\bibinfo {author} {\bibfnamefont {R.}~\bibnamefont
  {Feynman}}\ and\ \bibinfo {author} {\bibfnamefont {F.}~\bibnamefont
  {Vernon}},\ }\href
  {https://doi.org/https://doi.org/10.1016/0003-4916(63)90068-X} {\bibfield
  {journal} {\bibinfo  {journal} {Annals of Physics}\ }\textbf {\bibinfo
  {volume} {24}},\ \bibinfo {pages} {118} (\bibinfo {year} {1963})}\BibitemShut
  {NoStop}%
\bibitem [{\citenamefont {Vernon}(1959)}]{Vernon:1959vsj}%
  \BibitemOpen
  \bibfield  {author} {\bibinfo {author} {\bibfnamefont {F.~L.}\ \bibnamefont
  {Vernon}},\ }\emph {\bibinfo {title} {{The theory of a general quantum system
  interacting with a linear dissipative system}}},\ \href
  {https://doi.org/10.7907/JNJ1-QZ83} {Ph.D. thesis},\ \bibinfo  {school}
  {Caltech} (\bibinfo {year} {1959})\BibitemShut {NoStop}%
\bibitem [{\citenamefont {Abac}\ \emph {et~al.}(2026)\citenamefont {Abac} \emph
  {et~al.}}]{LIGOScientific:2026fcf}%
  \BibitemOpen
  \bibfield  {author} {\bibinfo {author} {\bibfnamefont {A.~G.}\ \bibnamefont
  {Abac}} \emph {et~al.} (\bibinfo {collaboration} {LIGO Scientific, VIRGO,
  KAGRA}),\ }\href@noop {} {\  (\bibinfo {year} {2026})},\ \Eprint
  {https://arxiv.org/abs/2603.19020} {arXiv:2603.19020 [gr-qc]} \BibitemShut
  {NoStop}%
\bibitem [{\citenamefont {Colpi}\ \emph {et~al.}(2024)\citenamefont {Colpi}
  \emph {et~al.}}]{LISA:2024hlh}%
  \BibitemOpen
  \bibfield  {author} {\bibinfo {author} {\bibfnamefont {M.}~\bibnamefont
  {Colpi}} \emph {et~al.} (\bibinfo {collaboration} {LISA}),\ }\href@noop {} {\
   (\bibinfo {year} {2024})},\ \Eprint {https://arxiv.org/abs/2402.07571}
  {arXiv:2402.07571 [astro-ph.CO]} \BibitemShut {NoStop}%
\bibitem [{\citenamefont {Kanno}\ and\ \citenamefont {Soda}(2019)}]{kanno1}%
  \BibitemOpen
  \bibfield  {author} {\bibinfo {author} {\bibfnamefont {S.}~\bibnamefont
  {Kanno}}\ and\ \bibinfo {author} {\bibfnamefont {J.}~\bibnamefont {Soda}},\
  }\href {https://doi.org/10.1103/PhysRevD.99.084010} {\bibfield  {journal}
  {\bibinfo  {journal} {Phys. Rev. D}\ }\textbf {\bibinfo {volume} {99}},\
  \bibinfo {pages} {084010} (\bibinfo {year} {2019})}\BibitemShut {NoStop}%
\bibitem [{\citenamefont {Kanno}(2019)}]{kanno2}%
  \BibitemOpen
  \bibfield  {author} {\bibinfo {author} {\bibfnamefont {S.}~\bibnamefont
  {Kanno}},\ }\href {https://doi.org/10.1103/PhysRevD.100.123536} {\bibfield
  {journal} {\bibinfo  {journal} {Phys. Rev. D}\ }\textbf {\bibinfo {volume}
  {100}},\ \bibinfo {pages} {123536} (\bibinfo {year} {2019})}\BibitemShut
  {NoStop}%
\bibitem [{\citenamefont {Mukhanov}\ and\ \citenamefont
  {Winitzki}(2007)}]{Mukhanov}%
  \BibitemOpen
  \bibfield  {author} {\bibinfo {author} {\bibfnamefont {V.}~\bibnamefont
  {Mukhanov}}\ and\ \bibinfo {author} {\bibfnamefont {S.}~\bibnamefont
  {Winitzki}},\ }\href@noop {} {\emph {\bibinfo {title} {Introduction to
  Quantum Effects in Gravity}}}\ (\bibinfo  {publisher} {Cambridge University
  Press},\ \bibinfo {year} {2007})\BibitemShut {NoStop}%
\bibitem [{\citenamefont {Dorlis}\ \emph {et~al.}(2026)\citenamefont {Dorlis},
  \citenamefont {Mavromatos}, \citenamefont {Sarkar},\ and\ \citenamefont
  {Vlachos}}]{Dorlis:2025amf}%
  \BibitemOpen
  \bibfield  {author} {\bibinfo {author} {\bibfnamefont {P.}~\bibnamefont
  {Dorlis}}, \bibinfo {author} {\bibfnamefont {N.~E.}\ \bibnamefont
  {Mavromatos}}, \bibinfo {author} {\bibfnamefont {S.}~\bibnamefont {Sarkar}},\
  and\ \bibinfo {author} {\bibfnamefont {S.-N.}\ \bibnamefont {Vlachos}},\
  }\href {https://doi.org/10.1103/t822-86qp} {\bibfield  {journal} {\bibinfo
  {journal} {Phys. Rev. D}\ }\textbf {\bibinfo {volume} {113}},\ \bibinfo
  {pages} {026023} (\bibinfo {year} {2026})},\ \Eprint
  {https://arxiv.org/abs/2507.23475} {arXiv:2507.23475 [gr-qc]} \BibitemShut
  {NoStop}%
\bibitem [{\citenamefont {Dorlis}\ \emph {et~al.}(2025)\citenamefont {Dorlis},
  \citenamefont {Mavromatos}, \citenamefont {Sarkar},\ and\ \citenamefont
  {Vlachos}}]{Dorlis:2025zzz}%
  \BibitemOpen
  \bibfield  {author} {\bibinfo {author} {\bibfnamefont {P.}~\bibnamefont
  {Dorlis}}, \bibinfo {author} {\bibfnamefont {N.~E.}\ \bibnamefont
  {Mavromatos}}, \bibinfo {author} {\bibfnamefont {S.}~\bibnamefont {Sarkar}},\
  and\ \bibinfo {author} {\bibfnamefont {S.-N.}\ \bibnamefont {Vlachos}},\
  }\href {https://doi.org/10.1103/9crd-zj6l} {\bibfield  {journal} {\bibinfo
  {journal} {Phys. Rev. Lett.}\ }\textbf {\bibinfo {volume} {135}},\ \bibinfo
  {pages} {151501} (\bibinfo {year} {2025})},\ \Eprint
  {https://arxiv.org/abs/2507.01689} {arXiv:2507.01689 [gr-qc]} \BibitemShut
  {NoStop}%
\bibitem [{\citenamefont {Mavromatos}\ and\ \citenamefont
  {Sarkar}(2026)}]{Mavromatos:2026nin}%
  \BibitemOpen
  \bibfield  {author} {\bibinfo {author} {\bibfnamefont {N.~E.}\ \bibnamefont
  {Mavromatos}}\ and\ \bibinfo {author} {\bibfnamefont {S.}~\bibnamefont
  {Sarkar}},\ }\href@noop {} {\  (\bibinfo {year} {2026})},\ \Eprint
  {https://arxiv.org/abs/2606.24613} {arXiv:2606.24613 [gr-qc]} \BibitemShut
  {NoStop}%
\end{thebibliography}%
 
\end{document}